\documentclass[journal,doublecolumn,12pt]{IEEEtran}
\usepackage{epsfig}
\usepackage{cite}
\usepackage{comment}
\usepackage{balance}
\usepackage{algpseudocode} 
\usepackage{setspace}
\usepackage{amsmath}
\usepackage[table]{xcolor}
\usepackage{lipsum}
\usepackage{array}
\usepackage{tabularx} 
\usepackage{booktabs} 
\usepackage{enumitem}
\usepackage{siunitx}
\usepackage{amsmath,amssymb,amsfonts}
\usepackage{mathtools}
\usepackage{pgfplots}
\usetikzlibrary{arrows}
\usepackage{color}
\usepackage{arydshln}
\usepackage{url}
\usepackage{soul}
\usepackage{csquotes}
\usepackage[compact]{titlesec}
\usepackage{graphicx}
\usepackage{svg}
\usepackage{epstopdf}
\usepackage[nolist]{acronym}
\usepackage{subcaption}
\usepackage[labelsep=period]{caption}

\usepackage{orcidlink}
\usepackage[capitalise]{cleveref}

\Crefname{equation}{Eq.\!}{Eqs.\!}
\Crefname{figure}{Fig.\!}{Figs.\!}
\Crefname{tabular}{Tab.\!}{Tabs.\!}
\Crefname{section}{Section\!}{Sections.\!}

\pgfplotsset{compat=1.18} 

\usepackage[ruled,vlined,commentsnumbered,linesnumbered,lined,boxed]{algorithm2e}
\SetInd{0em}{1em} 
\usepackage{algorithmicx}
\usepackage{algpseudocode}
\usepackage{cuted}
\usepackage{pgfplotstable}
\usepackage[utf8]{inputenc}

\definecolor{gray15}{rgb}{0.95, 0.95, 0.95}
\definecolor{mycolor1}{rgb}{0.494, 0.184, 0.556}
\definecolor{mycolor2}{rgb}{0.466, 0.674, 0.188}
\definecolor{mycolor3}{rgb}{0.301, 0.745, 0.933}
\definecolor{mycolor4}{rgb}{0.635, 0.078, 0.184}
\definecolor{mycolor5}{rgb}{0.000, 0.447, 0.741}

\SetAlgoNlRelativeSize{-1} 
\titlespacing{\section}{0pt}{*0.3}{*0.3}
\titlespacing{\subsection}{0pt}{*0.3}{*0.2}

\begin{document}

\bstctlcite{IEEEexample:BSTcontrol}
\newcommand{\norm}[1]{\left\lVert#1\right\rVert}
\newcommand{\normo}[1]{{\left\lVert#1\right\rVert}_{0}}
\def\pvl{\boldsymbol{\Gamma}[\ell]}
\def\pv{\boldsymbol{\Gamma}}
\def\pvu{\boldsymbol{\Gamma}_{u}[\ell]}
\def\pvun{\boldsymbol{\Gamma}_{u,n}}

\def\SAR{\mathrm{SAR}}
\def\Thetac{\boldsymbol{\Theta}}
\def\Hone{\mathbf{H}_{1}}
\def\Htwo{\mathbf{H}_{2}}
\def\Hthree{\mathbf{H}_{3}}
\def\x{\mathbf{x}}
\def\n{\mathbf{n}}
\def\u{\mathbf{u}}
\def\EI{I}

\def\mr{M_{\textrm{r}}}
\def\mt{M_{\textrm{t}}}
\def\nc{N_{\textrm{c}}}
\def\hd{\mathbf{H}^{\text{d}}}

\def\daggerr{\text{H}}



\def\sign{\textrm{sign}}                                              
\def\erf{\textrm{erf}}
\def\erfc{\textrm{erfc}}

\begin{acronym}
\acro{5G-NR}{5G New Radio}
\acro{3GPP}{3rd Generation Partnership Project}
\acro{AC}{address coding}
\acro{ACF}{autocorrelation function}
\acro{ACR}{autocorrelation receiver}
\acro{ADC}{analog-to-digital converter}
\acrodef{aic}[AIC]{Analog-to-Information Converter}     
\acro{AIC}[AIC]{Akaike information criterion}
\acro{aric}[ARIC]{asymmetric restricted isometry constant}
\acro{arip}[ARIP]{asymmetric restricted isometry property}
\acro{ARIS}{aerial reconfigurable intelligent surface}

\acro{ARQ}{automatic repeat request}
\acro{AUB}{asymptotic union bound}
\acrodef{awgn}[AWGN]{Additive White Gaussian Noise}     
\acro{AWGN}{additive white Gaussian noise}
\acro{APSK}[PSK]{asymmetric PSK} 
\acro{AO}{alternate optimization}
\acro{IoT}{Internet of Things}
\acro{IoE}{Internet of Everything}
\acro{ZF}{Zero-Forcing}

\acro{mmW}{Millimeter Waves}

\acro{waric}[AWRICs]{asymmetric weak restricted isometry constants}
\acro{warip}[AWRIP]{asymmetric weak restricted isometry property}
\acro{BCH}{Bose, Chaudhuri, and Hocquenghem}        
\acro{BCHC}[BCHSC]{BCH based source coding}
\acro{BEP}{bit error probability}
\acro{BFC}{block fading channel}
\acro{BG}[BG]{Bernoulli-Gaussian}
\acro{BGG}{Bernoulli-Generalized Gaussian}
\acro{BPAM}{binary pulse amplitude modulation}
\acro{BPDN}{Basis Pursuit Denoising}
\acro{BPPM}{binary pulse position modulation}
\acro{BPSK}{binary phase shift keying}
\acro{BPZF}{bandpass zonal filter}
\acro{BSC}{binary symmetric channels}              
\acro{BU}[BU]{Bernoulli-uniform}
\acro{BER}{bit error rate}
\acro{BS}{base station}
\acro{EI}{exposure index}
\acro{MU-MIMO}{multi-user multiple-input-multiple-output}
\acro{MU-SIMO}{multi-user single-input-multiple-output}

\acro{CP}{Cyclic Prefix}
\acrodef{cdf}[CDF]{cumulative distribution function}   
\acro{CDF}{cumulative distribution function}
\acrodef{c.d.f.}[CDF]{cumulative distribution function}
\acro{CCDF}{complementary cumulative distribution function}
\acrodef{ccdf}[CCDF]{complementary CDF}       
\acrodef{c.c.d.f.}[CCDF]{complementary cumulative distribution function}
\acro{CD}{cooperative diversity}
\acro{CDMA}{Code Division Multiple Access}
\acro{ch.f.}{characteristic function}
\acro{CIR}{channel impulse response}
\acro{cosamp}[CoSaMP]{compressive sampling matching pursuit}
\acro{CR}{cognitive radio}
\acro{cs}[CS]{compressed sensing}                  
\acrodef{cscapital}[CS]{Compressed sensing} 
\acrodef{CS}[CS]{compressed sensing}
\acro{CSI}{channel state information}
\acro{CCSDS}{consultative committee for space data systems}
\acro{CC}{convolutional coding}
\acro{Covid19}[COVID-19]{Coronavirus disease}

\acro{DAA}{detect and avoid}
\acro{DAB}{digital audio broadcasting}
\acro{DCT}{discrete cosine transform}
\acro{dft}[DFT]{discrete Fourier transform}
\acro{DR}{distortion-rate}
\acro{DS}{direct sequence}
\acro{DS-SS}{direct-sequence spread-spectrum}
\acro{DTR}{differential transmitted-reference}
\acro{DVB-H}{digital video broadcasting\,--\,handheld}
\acro{DVB-T}{digital video broadcasting\,--\,terrestrial}
\acrodef{DL}{downlink}
\acro{DSSS}{Direct Sequence Spread Spectrum}
\acro{DFT-s-OFDM}{Discrete Fourier Transform-spread-Orthogonal Frequency Division Multiplexing}
\acro{DAS}{distributed antenna system}
\acro{DNA}{Deoxyribonucleic Acid}

\acro{EC}{European Commission}
\acro{EED}[EED]{exact eigenvalues distribution}
\acro{EIRP}{Equivalent Isotropically Radiated Power}
\acro{ELP}{equivalent low-pass}
\acro{eMBB}{Enhanced Mobile Broadband}
\acro{EMF}{electromagnetic fields}
\acro{EM}{electromagnetic}
\acro{EU}{European union}

\acro{FC}[FC]{fusion center}
\acro{FCC}{Federal Communications Commission}
\acro{FEC}{forward error correction}
\acro{FFT}{fast Fourier transform}
\acro{FH}{frequency-hopping}
\acro{FH-SS}{frequency-hopping spread-spectrum}
\acrodef{FS}{Frame synchronization}
\acro{FSsmall}[FS]{frame synchronization}  
\acro{FDMA}{Frequency Division Multiple Access}

\acro{GA}{Gaussian approximation}
\acro{GF}{Galois field }
\acro{GG}{Generalized-Gaussian}
\acro{GIC}[GIC]{generalized information criterion}
\acro{GLRT}{generalized likelihood ratio test}
\acro{GPS}{Global Positioning System}
\acro{GMSK}{Gaussian minimum shift keying}
\acro{GSMA}{Global System for Mobile communications Association}
\acro{GR}{Gaussian randomization}
\acro{HAP}{high altitude platform}

\acro{IDR}{information distortion-rate}
\acro{IFFT}{inverse fast Fourier transform}
\acro{iht}[IHT]{iterative hard thresholding}
\acro{i.i.d.}{independent, identically distributed}
\acro{IoT}{Internet of Things}                      
\acro{IR}{impulse radio}
\acro{lric}[LRIC]{lower restricted isometry constant}
\acro{lrict}[LRICt]{lower restricted isometry constant threshold}
\acro{ISI}{intersymbol interference}
\acro{ITU}{International Telecommunication Union}
\acro{ICNIRP}{International Commission on Non-Ionizing Radiation Protection}
\acro{IEEE}{Institute of Electrical and Electronics Engineers}
\acro{ICES}{IEEE international committee on electromagnetic safety}
\acro{IEC}{International Electrotechnical Commission}
\acro{IARC}{International Agency on Research on Cancer}
\acro{IS-95}{Interim Standard 95}
\acro{i.i.d}{independent and identically distributed}

\acro{LEO}{low earth orbit}
\acro{LF}{likelihood function}
\acro{LLF}{log-likelihood function}
\acro{LLR}{log-likelihood ratio}
\acro{LLRT}{log-likelihood ratio test}
\acro{LOS}{Line-of-Sight}
\acro{LRT}{likelihood ratio test}
\acro{wlric}[LWRIC]{lower weak restricted isometry constant}
\acro{wlrict}[LWRICt]{LWRIC threshold}
\acro{LPWAN}{low power wide area network}
\acro{LoRaWAN}{Low power long Range Wide Area Network}
\acro{NLOS}{non-line-of-sight}

\acro{MB}{multiband}
\acro{MC}{multicarrier}
\acro{MDS}{mixed distributed source}
\acro{MF}{matched filter}
\acro{m.g.f.}{moment generating function}
\acro{MI}{mutual information}
\acro{MIMO}{multiple-input multiple-output}
\acro{MISO}{multiple-input single-output}
\acrodef{maxs}[MJSO]{maximum joint support cardinality}                       
\acro{ML}[ML]{maximum likelihood}
\acro{MMSE}{minimum mean-square error}
\acro{MMV}{multiple measurement vectors}
\acrodef{MOS}{model order selection}
\acro{M-PSK}[${M}$-PSK]{$M$-ary phase shift keying}                       
\acro{M-APSK}[${M}$-PSK]{$M$-ary asymmetric PSK} 

\acro{M-QAM}[$M$-QAM]{$M$-ary quadrature amplitude modulation}
\acro{MRC}{maximal ratio combiner}                  
\acro{maxs}[MSO]{maximum sparsity order}                                      
\acro{M2M}{machine to machine}                                                
\acro{MUI}{multi-user interference}
\acro{mMTC}{massive Machine Type Communications}      
\acro{mm-Wave}{millimeter-wave}
\acro{MP}{mobile phone}
\acro{MPE}{maximum permissible exposure}
\acro{MAC}{media access control}
\acro{MUMIMO}{multi-user  \ac{MIMO}}

\acro{NLoS}{non line-of-sight}
\acro{NB}{narrowband}
\acro{NBI}{narrowband interference}
\acro{NLA}{nonlinear sparse approximation}
\acro{NTIA}{National Telecommunications and Information Administration}
\acro{NTP}{National Toxicology Program}
\acro{NHS}{National Health Service}

\acro{OC}{optimum combining}                       \acro{QCQP}  {quadratically constrained quadratic problem}
\acro{OC}{optimum combining}
\acro{ODE}{operational distortion-energy}
\acro{ODR}{operational distortion-rate}
\acro{OFDM}{orthogonal frequency-division multiplexing}
\acro{omp}[OMP]{orthogonal matching pursuit}
\acro{OSMP}[OSMP]{orthogonal subspace matching pursuit}
\acro{OQAM}{offset quadrature amplitude modulation}
\acro{OQPSK}{offset QPSK}
\acro{OFDMA}{Orthogonal Frequency-division Multiple Access}
\acro{OPEX}{Operating Expenditures}
\acro{OQPSK/PM}{OQPSK with phase modulation}

\acro{PAM}{pulse amplitude modulation}
\acro{PAR}{peak-to-average ratio}
\acrodef{pdf}[PDF]{probability density function}                      
\acro{PDF}{probability density function}
\acrodef{p.d.f.}[PDF]{probability distribution function}
\acro{PDP}{power dispersion profile}
\acro{PMF}{probability mass function}                             
\acrodef{p.m.f.}[PMF]{probability mass function}
\acro{PN}{pseudo-noise}
\acro{PPM}{pulse position modulation}
\acro{PRake}{Partial Rake}
\acro{PSD}{power spectral density}
\acro{PSEP}{pairwise synchronization error probability}
\acro{PSK}{phase shift keying}
\acro{PD}{power density}
\acro{8-PSK}[$8$-PSK]{$8$-phase shift keying}

\acro{FSK}{frequency shift keying}

\acro{QAM}{Quadrature Amplitude Modulation}
\acro{QPSK}{quadrature phase shift keying}
\acro{OQPSK/PM}{OQPSK with phase modulator }

\acro{RE}{resource element}
\acro{RD}[RD]{raw data}
\acro{RDL}{"random data limit"}
\acro{ric}[RIC]{restricted isometry constant}
\acro{rict}[RICt]{restricted isometry constant threshold}
\acro{rip}[RIP]{restricted isometry property}
\acro{ROC}{receiver operating characteristic}
\acro{rq}[RQ]{Raleigh quotient}
\acro{RS}[RS]{Reed-Solomon}
\acro{RSC}[RSSC]{RS based source coding}
\acro{r.v.}{random variable}                               
\acro{R.V.}{random vector}
\acro{RMS}{root mean square}
\acro{RFR}{radiofrequency radiation}
\acro{RIS}{reconfigurable intelligent surface}
\acro{RNA}{RiboNucleic Acid}

\acro{LoS}{line-of-sight}

\acro{SA}[SA-Music]{subspace-augmented MUSIC with OSMP}
\acro{SAR}{specific absorption rate}
\acro{SCBSES}[SCBSES]{Source Compression Based Syndrome Encoding Scheme}
\acro{SCM}{sample covariance matrix}
\acro{SDR}{semi-definite relaxation}
\acro{SDP}{semi-definite program}
\acro{SEP}{symbol error probability}
\acro{SG}[SG]{sparse-land Gaussian model}
\acro{SIMO}{single-input multiple-output}
\acro{SINR}{signal-to-interference plus noise ratio}
\acro{SIR}{signal-to-interference ratio}
\acro{SISO}{single-input single-output}
\acro{SMV}{single measurement vector}
\acro{SNR}[\textrm{SNR}]{signal-to-noise ratio} 
\acro{sp}[SP]{subspace pursuit}
\acro{SS}{spread spectrum}
\acro{SW}{sync word}
\acro{SAR}{specific absorption rate}
\acro{SSB}{synchronization signal block}
\acro{SCA}{successive convex approximation}

\acro{TH}{time-hopping}
\acro{ToA}{time-of-arrival}
\acro{TR}{transmitted-reference}
\acro{TW}{Tracy-Widom}
\acro{TWDT}{TW Distribution Tail}
\acro{TCM}{trellis coded modulation}
\acro{TDD}{time-division duplexing}
\acro{TDMA}{Time Division Multiple Access}

\acro{UAV}{unmanned aerial vehicle}
\acro{uric}[URIC]{upper restricted isometry constant}
\acro{urict}[URICt]{upper restricted isometry constant threshold}
\acro{UWB}{ultrawide band}
\acro{UWBcap}[UWB]{Ultrawide band}   
\acro{URLLC}{Ultra Reliable Low Latency Communications}
         
\acro{wuric}[UWRIC]{upper weak restricted isometry constant}
\acro{wurict}[UWRICt]{UWRIC threshold}                
\acro{UE}{user equipment}
\acrodef{UL}{uplink}

\acro{ULA}{uniform linear array}

\acro{WiM}[WiM]{weigh-in-motion}
\acro{WLAN}{wireless local area network}
\acro{wm}[WM]{Wishart matrix}                               
\acroplural{wm}[WM]{Wishart matrices}
\acro{WMAN}{wireless metropolitan area network}
\acro{WPAN}{wireless personal area network}
\acro{wric}[WRIC]{weak restricted isometry constant}
\acro{wrict}[WRICt]{weak restricted isometry constant thresholds}
\acro{wrip}[WRIP]{weak restricted isometry property}
\acro{WSN}{wireless sensor network}                        
\acro{WSS}{wide-sense stationary}
\acro{WHO}{World Health Organization}
\acro{Wi-Fi}{wireless fidelity}

\acro{sss}[SpaSoSEnc]{sparse source syndrome encoding}

\acro{VLC}{visible light communication}
\acro{VPN}{virtual private network} 
\acro{RF}{radio frequency}
\acro{FSO}{free space optics}
\acro{IoST}{Internet of space things}

\acro{GSM}{Global System for Mobile Communications}
\acro{2G}{second-generation cellular network}
\acro{3G}{third-generation cellular network}
\acro{4G}{fourth-generation cellular network}
\acro{5G}{fifth-generation}	
\acro{gNB}{next generation node B base station}
\acro{NR}{New Radio}
\acro{UMTS}{Universal Mobile Telecommunications Service}
\acro{LTE}{Long Term Evolution}

\acro{QoS}{quality of service}
\acro{KKT}{Karush–Kuhn–Tucker}

\acro{WPD}{wireless power transfer device}
\acro{w.r.t}{with respect to}
\end{acronym}

\title{EMF-Efficient MU-MIMO Networks: Harnessing Aerial RIS Technology}
\author{
    Mariem Chemingui \orcidlink{https://orcid.org/0000-0002-6394-4469}, 
    Ahmed Elzanaty \orcidlink{https://orcid.org/0000-0001-8854-8369},
    Rahim Tafazolli \orcidlink{https://orcid.org/0000-0002-6062-8639}\\Institute for Communication Systems (ICS), University of Surrey, UK 
  
}
	\maketitle
\begin{abstract}
The rollout of the \ac{5G} networks has raised some concerns about potential health effects from increased exposure to \ac{EMF}. To address these concerns, we design a novel \ac{EMF}-aware architecture for uplink communications. Specifically, we propose an \ac{ARIS} assisted multi-user \ac{MIMO} system, where the \ac{ARIS} features a \ac{RIS} panel mounted on an \ac{UAV}, offering a flexible and adaptive solution for reducing uplink EMF exposure. We formulate and solve a new problem to minimize the \ac{EMF} exposure by optimizing the system parameters, such as transmit beamforming, resource allocation, transmit power, \ac{ARIS} phase shifts, and \ac{ARIS} trajectory. Our numerical results demonstrate the effectiveness of \ac{EMF}-aware transmission scheme over the benchmark methods, achieving EMF reductions of over $30\%$ and $90\%$ compared to the fixed ARIS and non-ARIS schemes, respectively.
     \end{abstract}
 \begin{IEEEkeywords}
     \ac{5G}, electromagnetic field (EMF), EMF-aware design, multi-user MIMO, aerial reconfigurable intelligent surface (ARIS).
 \end{IEEEkeywords}
 \acresetall
\section{Introduction}
 Nowadays, the widespread deployment of \ac{5G} technology has raised concerns among certain segments of the population regarding the potential exposure to \ac{EMF} emitted by \ac{5G} networks \cite{nyberg20175g,32021}. 
 The potential for heightened \ac{EMF} exposure is contingent not only upon network architecture but also on the characteristics of the utilized devices. Recently, France has halted the sales of the iPhone 12 due to its higher radiation levels exceeding regulatory limits \cite{bbc}.
 
 Studies have indicated that \ac{EM} radiations can have both thermal and non-thermal effects on the human body \cite{aldrich1987electromagnetic,jamshed2019survey,elzanaty20215g}. The thermal effects occur when the body absorbs excessive \ac{EM} radiation, leading to the production of heat that can potentially damage tissues. However, the scientific community continues to engage in ongoing debates regarding the potential long-term non-thermal effects of \ac{EM} radiation on human health \cite{yinhui2019effect,national2018toxicology,ahlbom2008possible}.
 Nevertheless, certain investigations have documented discernible biological manifestations attributed to exposure to non-ionizing \ac{EMF} \cite{interphone2011acoustic,larjavaara2011location}. In this context, the International Agency for Research on Cancer (IARC) has categorized radio frequency radiation as \textquote{Possibly carcinogenic to humans} (Group 2B) \cite{iarc2013non}.
 
To address the public concern, regulatory agencies such as the Federal Communications Commission (FCC) \cite{FCC} and the International Commission on Non-Ionizing Radiation Protection (ICNIRP)\cite{ICNIRP}, have established guidelines to ensure that the \ac{EMF} exposure is quantified and limited on the users' side. However, \ac{EMF} exposure refers to the radiation exposure generated by the propagation of the \ac{EM} waves, which are usually emitted by wireless terminals such as \acp{BS} and \ac{UE}. Since \acp{UE} are much closer to the human body compared to \ac{BS}, in general, the uplink \ac{EMF} exposure from them is dominant, i.e., approximately ten times stronger than the downlink exposure \cite{ULDominance,castellanos2022multi,chiaraviglio2023dominance}. The dominance of uplink exposure is further supported by studies in large-scale cellular networks \cite{qin2023unveiling,gontier2023uplink}.

To evaluate \ac{EMF} exposure during \ac{UL} transmission, the standard \ac{SAR} metric, which measures the amount of electromagnetic power absorbed by human tissue per unit mass, is utilized \cite{SARCode2013,heliot2020exposure}. Consequently, many governments require that uplink EMF exposure, as assessed by \ac{SAR}, remains below specified thresholds. In response to this, precautionary measures should be taken while designing future cellular systems.
\subsection{Related Work}
In the \ac{5G} cellular systems, the prevalence of connected devices with multiple antennas has evoked \ac{EMF}-aware design for the uplink transmission. For example, authors in \cite{5723738} proposed a beamforming-based technique for minimizing the \ac{EMF} exposure in the uplink transmission design. In \cite{Fabian'spaper}, the authors provided a model to assess the individual exposure with multi-antenna user terminals operating in a multi-carrier environment by optimizing the relative phase differences and power allocation between the antenna ports. However, when considering higher frequency bands, the \ac{LoS} blockage needs to be accounted for, as it can significantly influence the distribution and intensity of electromagnetic fields and subsequent exposure levels, as the users may need to increase their transmit power to compensate for the high path-loss \cite{Elzanaty-Emf}.

Recently, due to the proposal of the controllable intelligent radio environments, the \ac{RIS} has been introduced to mitigate the effect of \ac{LoS} blockages in terrestrial and non-terrestrial networks \cite{ReviwerSuggI, ReviwerSuggII}. The \ac{RIS} is a passive reflector that superimposes the incident signal waves by adjusting the phase shifts and then reflects them in the appropriate directions \cite{famousRISpaper,networkCodingAhmed}. Consequently, optimizing the phase shifts enhances the channel, enabling a reduction in transmit power without compromising the data rate \cite{RISPowScale}. In \cite{Elzanaty-Emf}, phase shifts of the \ac{RIS} were designed to reduce the exposure for uplink multi-user \ac{SIMO} systems. Note that the fixed placement of the \ac{RIS} may constrain its ability to effectively mitigate network blockages and optimize exposure reduction in scenarios characterized by dynamic changes in signal propagation. A possible solution for mobility is considering \ac{UAV}-assisted networks \cite{7470933,8764406,ReviwerIIR3}. In a notable example, the work in \cite{TethredUAVAhmed} presented an approach involving tethered \acp{UAV} equipped with receive-only antennas, aiming to minimize \ac{EMF} exposure while ensuring high \ac{QoS} for users. However, using tethered \acp{UAV} introduces constraints on mobility, resulting in reduced degrees of freedom when designing the network architecture. Furthermore, the restricted mobility of the tethered \acp{UAV} may hinder their ability to adapt to dynamic propagation environments or to  effectively satisfy evolving user requirements. Nevertheless, the ongoing investigation into \ac{RIS} has led to the introduction of a novel architecture known as \ac{ARIS}. Distinguishing itself from traditional \ac{RIS}, the \ac{ARIS} can be installed on aerial platforms like \acp{UAV}, enabling the establishment of more robust \ac{LoS} connections \cite{ARIS1}. Moreover, the \ac{ARIS} exhibits the capability to dynamically adjust its position, allowing it to adapt to various propagation environments.

In this context, we propose a two-fold design aimed at minimizing \ac{EMF} exposure. Firstly, we introduce \ac{ARIS} to achieve finer control over transmit beamforming, thereby reducing the coupling between antennas and users' bodies and minimizing the induced \ac{SAR}. Secondly, we adopt a power control scheme that reduces transmit power while ensuring \ac{QoS} for all network users. By optimizing the propagation environment through the design of ARIS phases and trajectory, we enhance uplink communication with \ac{LoS} links, thereby reducing the required transmit power. Additionally, resource and power allocation are managed to meet \ac{QoS} requirements effectively. Our novel architecture, incorporating \ac{ARIS}, dynamically adjusts propagation environment to minimize population EMF exposure, offering a cost-effective and sustainable solution for reducing radiation in various scenarios characterized by a low probability of \ac{LoS} links. To the best of our knowledge, \ac{EMF}-efficient architectures and associated algorithms involving \ac{ARIS} have not been previously discussed in the literature.
\subsection{Contributions}
{In this paper, we propose a novel architecture for \ac{EMF}-aware \ac{MU-MIMO} cellular systems, where the exposure is further reduced beyond the standards as precautionary measure.} More precisely, we consider \ac{ARIS} to assist uplink communications, aiming to reduce the overall \ac{EMF} exposure for the population while assuring the required transmit data rates for users. In fact, having a \ac{RIS} carried on a mobile \ac{UAV} can provide additional degrees of freedom in network design, bringing several benefits such as providing \textit{(i)} flexible deployment for \ac{RIS}, especially in dynamic networks; \textit{(ii)}  an indirect \ac{LoS}, with high probability, through the \ac{RIS} due to the relatively high altitude of the \ac{UAV}; and \textit{(iii)} \ac{EMF}-efficient solution  by using lower transmit power while maintaining \ac{QoS} standards, benefiting from improved \ac{SNR} due to the large number of \ac{RIS} elements.

The main contributions of this paper are summarized as follows.
\begin{itemize}
    \item We propose a novel \ac{EMF}-aware design where the transmit beamforming parameters, the phase-shift matrix of the \ac{RIS}, the \ac{RE} allocation vector, the transmit power of the multi-antenna \acp{UE}, and the \ac{ARIS} trajectory are optimized.
    \item In contrast to \cite{Elzanaty-Emf}, our work considers mobile \ac{ARIS}, where its trajectory is designed to minimize \ac{EMF} exposure.
    \item A multi-variable optimization is formulated to reduce the average uplink exposure of the population while maintaining the \ac{QoS} of all the users and taking into consideration the maximum transmit power of users, the maximum distance an \ac{ARIS} can travel, and unique channel allocation for users.
    \item Solving optimization problem in its original form by jointly optimizing all the variables is complex due to its nature as a nonlinear fractional mixed-integer programming problem. To simplify the problem, we propose an \ac{AO}-based algorithm where optimization variables are optimized alternately. Moreover, we assess the performance of the proposed scheme against other architectures, e.g., without \ac{RIS} \cite{Fabian'spaper}, with fixed \ac{RIS} \cite{Elzanaty-Emf}, and non-optimized \ac{RIS} phases. We then conduct a complexity analysis of our algorithm and compare it with benchmark schemes.
\end{itemize}
The rest of the paper is structured as follows. Section~\ref{sec:SignalModel} explains the considered system setup and signal model. Section~\ref{sec:PbFormulation} presents the problem formulation for minimizing the population's exposure. Subsequently, we present the proposed solution to the optimization problem. A comprehensive analysis of the algorithm's complexity is then provided. Simulation results are presented in Section~\ref{sec:numericalresults}. Lastly, Section~\ref{Sec:Conclusion} serves as the conclusion of the paper, encapsulating final remarks and interpretations.
\subsection{Notation and Organization}
This paper uses the following notations. Lower-case (e.g., $x$),  bold lower-case letter (e.g., $\boldsymbol{x}$), and bold upper-case letters (e.g., $\boldsymbol{X}$) denote scalars, vectors, and matrices, respectively.
We use $\mathbb{R}$ to denote set of real numbers and $\mathbb{C}$ to denote the set of complex numbers while $j$ represents the imaginary unit and $\textrm{arg}(\cdot)$ denotes the argument of a complex number.
For a matrix $\mathbf{a}$, $\mathbf{a}^\top$ refers to its transpose, and $\mathbf{a}^\daggerr$ refers to its conjugate transpose. The operator $\mathrm{diag}\left(\cdot\right)$ denotes the construction of a diagonal matrix based on an input vector, while $\mathrm{tr}(\cdot)$ denotes the trace of a matrix, and the operator $\|\cdot\|$ represents the $\ell_2$ norm. The notation $\mathbf{A}\succeq 0$ indicates that the matrix $\mathbf{A}$ is positive semidefinite. The optimal solution is denoted by $(.)^\ast$.
\begin{table}[t!]
\small
\centering 
\caption{\\  {\textsc{Summary of Notations}}}
\renewcommand{\arraystretch}{0.5} 
\begin{tabularx}{0.9\columnwidth}{l X} 
    \toprule
     {\textbf{Notation}} &  {\textbf{Description}} \\ 
    \midrule 
    \midrule
     {$\mathbf{G}$} &  {Channel matrix between users and ARIS}  \\
     {$\mathbf{H}$} &  {Channel matrix between ARIS and BS}\\
     {$\mathbf{H}^\text{d}$} &  {Direct channel matrix between users and BS} \\
     {$\mathbf{H}$} &  {Overall channel matrix between users and BS} \\
     {$\boldsymbol{\Theta}$} &  {Phase-shift matrix of the RIS} \\
     {$\mathbf{f}$} &  {Transmit beamforming vector} \\
     {$\mathbf{y}_{u}$} &  {Received signal at the BS from user $u$} \\
     {$\mathbf{s}$} &  {Transmitted uplink signal } \\
     {$\boldsymbol{\nu}$} &  {Additive white Gaussian noise at the BS} \\
     {$\mathbf{p}$} &  {Transmit power matrix }\\
     {$\boldsymbol{\delta}$} &  {Resource allocation matrix}\\
     {$(\boldsymbol{\alpha}, \boldsymbol{\beta})$} &  {Beamforming parameters} \\
     {$\mathbf{Q}$} &  {ARIS trajectory matrix} \\
     {$\gamma_{u_{\text{R}}},\gamma_{u_{\text{D}}}$} &  {Sine of arrival and departure angles at the ARIS}\\ 
     {$\eta_\text{A},\eta_\text{D}$} &  {Sine of arrival and departure angles at the BS}\\
     {$d_{u\text{R}},d_{\text{RB}}$}& {Distance between user and ARIS and between ARIS and BS respectively}\\
    \bottomrule
\end{tabularx}
\label{tab1}
\end{table}
\section{System Model and Problem Formulation}
\label{sec:SignalModel}
In this section, we present the system setup, the signal, and the \ac{EMF} exposure models. Then, we establish the problem formulation. 
\subsection{System Setup and Signal Model}
\begin{figure}[t]
    \begin{center}
\includegraphics[width =0.8\columnwidth]{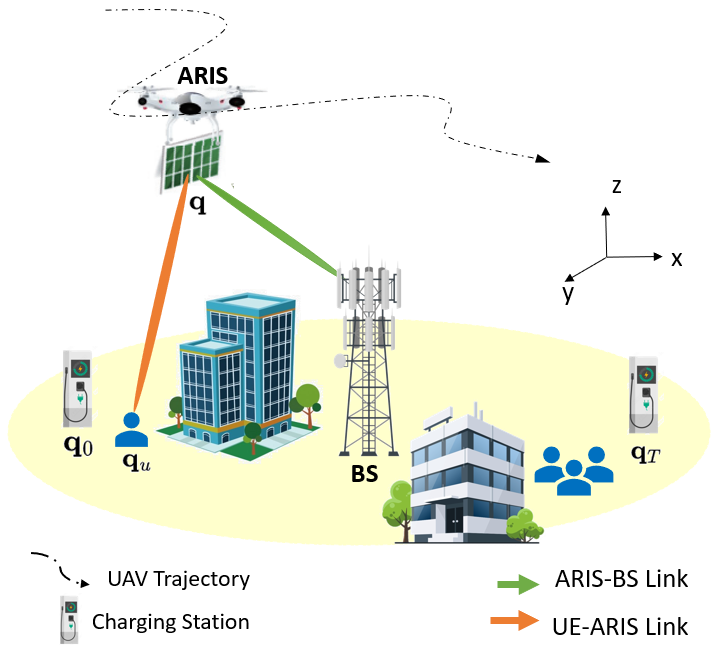}     \caption{ARIS-assisted network.}
    \label{fig:ARIS}
    \end{center}
\end{figure}

We consider an \ac{ARIS}-assisted \ac{MU-MIMO} network that consists of $U$ uplink users. Each \ac{UE} is equipped with $\mt$ antennas, while the \ac{BS} has $\mr$ receiving antennas. We assume all the communicating nodes are equipped with \acp{ULA}. In our scheme, the 
\ac{BS} is located at the center of the cell, i.e., $\mathbf{q}_{\text{B}}= (x_\text{B},y_\text{B},h_\text{B})=(0,0,h_{\text{B}})$, while users are randomly distributed inside the cell at $\mathbf{q}_u=(x_u,y_u,0), \forall u \in \mathcal{U}\triangleq  \{1,2,\cdots, U\}$. As depicted in Fig.~\ref{fig:ARIS}, the \ac{ARIS} is a \ac{RIS} deployed on a \ac{UAV} flying at a fixed altitude $h_0$. Through the navigation of the \ac{UAV} within the cell, the \ac{RIS} reflects the signals received from the transmitters to the \ac{BS} via $N$ passive elements. For tractability, the flying duration of the \ac{UAV} $T$ is discretized into $N_T$ time slots, each with duration $\Delta={T}/{N_T}$. Consequently, the \ac{UAV} trajectory is approximated by the sequence $\mathbf{Q}= \{\mathbf{q}[\ell], \forall \ell \in \mathcal{N_T}\}$, where $\mathbf{q}[\ell] =(x[\ell],y[\ell],h_0), \forall \ell \in \mathcal{N_T}\triangleq\{1,2,\dots,N_T\}$. In addition, the maximum distance of one time slot is $D_{\max}= \Delta\,V_{\text{max}}$, where $V_\text{max}$ denotes the maximum speed. The \ac{UAV} flies from an initial location $\mathbf{q}_0= (x_0,y_0,h_0)$ to a final location $\mathbf{q}_T= (x_T,y_T,h_0)$ during the considered flying duration before needing to be recharged.

We consider a multi-carrier system  with $\nc$ subchannels. Hence, the channel between user $u$ and the \ac{ARIS} through the subcarrier $n$, $\mathbf{G}_{u,n}[\ell] \in \mathbb{C}^{N \times \mt}$, can be modeled as a Rician channel, as the paths involving \ac{ARIS} usually have a \ac{LoS} component due to the high altitude of the \ac{UAV}. At time slot $\ell$, the channel can be expressed as 
\begin{equation} \small
\mathbf{G}_{u,n}[\ell]=\sqrt{\rho\, d_{u\text{R}}^{-\kappa_1}[\ell]} \bar{\mathbf{G}}_{u,n}[\ell],
\end{equation}
where 
\begin{equation} \small
   \bar{\mathbf{G}}_{u,n}[\ell]=  \sqrt{\frac{K_1}{K_1+1}} \mathbf{G}_{u,n}^{\text{LoS}}[\ell]+\sqrt{\frac{1}{K_1+1}} \mathbf{G}_{u,n}^{\text{NLoS}}[\ell].
\end{equation}
Here, $\rho$ accounts for the received power from this path at the reference distance (1m), $\kappa_1$ is the path-loss exponent, $K_1$ denotes the Rician factor of the channel, and $d_{u\text{R}}[\ell]=\|\mathbf{q}_u-\mathbf{q}[\ell]\|$ is the distance between the user $u$  and the \ac{RIS}. Meanwhile, the \ac{LoS} and \ac{NLoS} channel components are defined as $\mathbf{G}_{u,n}^{\text{LoS}}[\ell]$ and $\mathbf{G}_{u,n}^{\text{NLoS}}[\ell]$, where all elements of $\mathbf{G}_{u,n}^{\text{NLoS}}[\ell]$  are \ac{i.i.d} complex Gaussian random variables with zero mean
and unit variance. However, the matrix $\mathbf{G}_{u,n}^{\text{LoS}}[\ell]$ is expressed as 
\begin{equation} \small
\mathbf{G}_{u,n}^{\text{LoS}}[\ell] =\mathbf{a}(N,\gamma_{u_{\text{R}}}[\ell])~\mathbf{a}^\daggerr(\mt,\gamma_{u_{\text{D}}}[\ell]),
\end{equation}
where $\mathbf{a}$ denotes the steering vector, $\gamma_{u_{\text{R}}} [\ell] = \frac{y_u -y[\ell]}{d_{u\text{R}}[\ell]}$ and $\gamma_{u_{\text{D}}}[\ell] = \frac{y_u -y[\ell]}{d_{u\text{R}}[\ell]}$ are the sine of the arrival and departure angles at the \ac{ARIS}, respectively. 
 The steering vector $\mathbf{a}$ is a function of the number of antennas $M$ and the sine of the angle between the transmitter and the receiver $\gamma$, i.e.,
\begin{equation} \small
    \mathbf{a}(M,\gamma)\triangleq [1,e^{-j\frac{2\pi}{\lambda} d\, \gamma},\dots,e^{-j\frac{2\pi}{\lambda} d (M - 1)\gamma }]^\top,
\end{equation}
where $\lambda$ is the wavelength and $d$ the inter-antenna spacing \cite{RIS1}. Similarly, we define the channel between the \ac{RIS} and \ac{BS} through the \ac{RE} $n$ as follows
\begin{equation} \small
\mathbf{H}_{\mathrm{n}}[\ell]=\sqrt{\rho~d_{\text{RB}}^{-\kappa_2}[\ell]} \ \bar{\mathbf{H}}_{n}[\ell],
\end{equation}
where 
\begin{equation} \small
\bar{\mathbf{H}}_{n}[\ell]=\sqrt{\frac{K_2}{K_2+1}}~ \mathbf{H}_{n}^\text{LoS}[\ell]+\sqrt{\frac{1}{K_2+1}}~ \mathbf{H}_{n}^\text{NLoS}[\ell],
\end{equation}
with $\kappa_2$ as the path-loss exponent of the channel, $K_2$ the Rician factor and $d_{\text{RB}}[\ell]=\|q_{\text{B}}-q[\ell]\|$ the distance between the \ac{ARIS} and the \ac{BS} at $\ell$ time. The matrix $\bar{\mathbf{H}}_{\mathrm{n}}[\ell] = \mathbf{a}(\mr,\eta_{\text{D}}[\ell])~\mathbf{a}(N,\eta_{\text{A}}[\ell])^\daggerr,$ represents the \ac{LoS} component of the channel, depending on the sine of the arrival angle $\eta_{\text{A}}[\ell] = (x_\text{B}- x[\ell])/d_{\text{RB}}[\ell]$, and the departure angle $\eta_{\text{D}}[\ell] = (x_\text{B}- x[\ell])/d_{\text{RB}}[\ell]$. Each element of the \ac{NLoS} component $\tilde{\mathbf{H}}_{\mathrm{n}}[\ell]$ is \ac{i.i.d} complex Gaussian distribution with zero mean and unit variance. For the direct link between the users and the \ac{BS}, we consider a scenario where  {there is no direct \ac{LoS},} i.e., the Rician factor goes to zero. Therefore, we define $\hd_{u,n} \in \mathbb{C}^{\mr \times \mt}$ as the \ac{NLoS} channel between the user $u$ and the \ac{BS} via the $n^\text{th}$ subcarrier, which can be modeled as Rayleigh fading channel with variance $(\rho_1 d_{u\text{B}}^{-\kappa})$. 

The phase-shift matrix of the \ac{RIS} is defined as $\boldsymbol{\Theta}[\ell] = \mathrm{diag}\left(\boldsymbol{\theta}[\ell]\right)$, where $\boldsymbol{\theta}[\ell] \triangleq [e^{j \theta_1[\ell]},\!\dots\!, e^{j \theta_N[\ell]}]$ and $\theta_i[\ell]$ is the induced shift by the \ac{RIS} element $i$, $i\in \ \mathcal{N} \triangleq \{1,2,\cdots,N\}$.
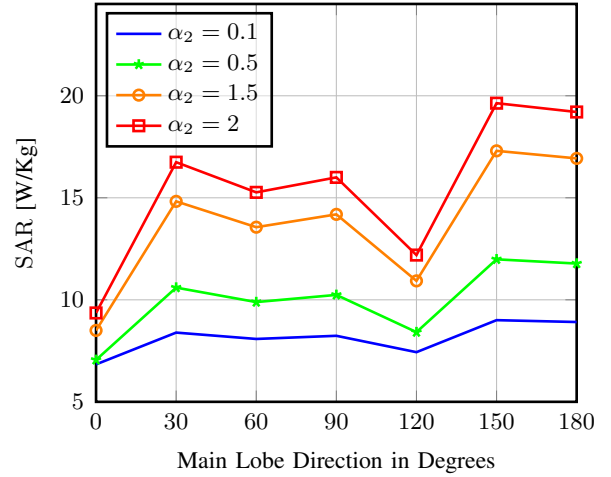
\begin{figure}[t!]
    \centering
    \pgfplotsset{every axis/.append style={
    font=\footnotesize,
    line width=1pt,
    legend style={
    fill opacity=0.5, %
        draw opacity=1,  
        text opacity=1,  
        fill=white,font=\footnotesize, at={(0.02,0.98)}, anchor=north west}, 
    legend cell align=left
}}
\pgfplotsset{compat=1.18}
   \begin{tikzpicture}
    \begin{axis}[
        xlabel near ticks,
        ylabel near ticks,
        grid=major,
        xlabel={Main Lobe Direction in Degrees},
        ylabel={$\SAR$ [W/Kg]},
         width=0.9\columnwidth,
        legend entries={$\alpha_2=0.1$, $\alpha_2=0.5$, $\alpha_2=1.5$,$\alpha_2=2\phantom{.0}$},
         xtick ={0,30,60,90,120,150,180},
         xmax = 180, xmin=0,
        ymin=5, ymax=24.5,
    ]
    \addplot[blue] table{Figures/SARFigure/SARPlotforalpha0.1.dat};
    \addplot[green, mark=star] table{Figures/SARFigure/SARPlotforalpha0.5.dat};
    \addplot[orange, mark=o] table{Figures/SARFigure/SARPlotforalpha1.5.dat};
    \addplot[red, mark=square] table{Figures/SARFigure/SARPlotforalpha2.dat};
    \end{axis}
    \end{tikzpicture}
    \caption{$\SAR$ as a function of the main lobe direction.}
    \label{fig:exp/var}
\end{figure}
The signal transmitted from the user $u$ to the \ac{BS} via the subcarrier $n$ at $\ell$ time is expressed as $\mathbf{s}_{u,n}[\ell] =\sqrt{p_{u,n}[\ell]} \mathbf{f}_{u,n}[\ell] x_{u,n}[\ell]$, where $x_{u,n}[\ell]$ is power-normalized transmitted information symbol and $\mathbf{f}_{u,n}[\ell] \in \mathbb{C}^{\mt \times 1} $ is the beamforming vector at user $u$. Therefore, the received signal at the \ac{BS} from user $u$ at time slot $\ell$ can be written as 
\begin{align}
    \small \boldsymbol{y}_u[\ell] = \sum_{n=1 }^{\nc}\delta_{u, n}[\ell]\: \mathbf{H}_{u,n}[\ell]\:\mathbf{s}_{u,n}[\ell] +\boldsymbol{\nu}[\ell], \forall\, u \in \mathcal{U},
\end{align}
where $\mathbf{H}_{u,n}[\ell] = \mathbf{H}_{n}[\ell]\, \boldsymbol{\Theta}[\ell]\, \mathbf{G}_{u,n}[\ell] +\mathbf{H}_{u,n}^\text{d}[\ell]$ denotes the overall channel matrix, $\delta_{u,n}[\ell]$ is a binary \ac{RE} allocation indicator, i.e., $\delta_{u,n}[\ell]=1$ if the $n^{\text{th}}$ RE is assigned to user $u$ and zero otherwise, and $\boldsymbol{\nu}$ represents the \ac{AWGN} at the \ac{BS} antennas with elements drawn from a zero-mean complex Gaussian distribution with variance $\sigma^2$. 

In the considered scenario, the achievable rate of the user $u$ using \ac{RE} $n$ is given by 
\begin{equation} \small
\small r_{u,n}(\pvl)=\omega \:\delta_{u, n}[\ell] \log_{2}\left(1+\frac{p_{u, n}[\ell]}{\sigma^{2}} \gamma(\pvun[\ell]) \right),
\label{rateequn}
\end{equation}
where $\omega$ denotes the bandwidth of each \ac{RE} and $\pv \triangleq \left(\boldsymbol{\delta}, \mathbf{P}, \boldsymbol{\alpha},\boldsymbol{\beta},\boldsymbol{\theta},\mathbf{Q}\right)$ represents all system parameters over all subcarriers for all time slots. Specifically, $\boldsymbol{\delta}$ represents the resource allocation for all users, $\mathbf{P}$ denotes the transmit power allocation matrix for all users, $(\boldsymbol{\alpha},\boldsymbol{\beta})$ is the beamforming parameters for all users, and $\boldsymbol{\theta}$ is the RIS phase-shift matrix over all time frame. Thus, the channel gain is given by
\begin{equation} \small
\small 
\begin{aligned}
    &  \gamma(\pvun[\ell]) = \mathbf{f}_{u, n}^\daggerr[\ell]\ \mathbf{K}_{u,n}[\ell] \ \mathbf{f}_{u, n}[\ell] \\
    &= \sum_{i=1}^{\mt} \alpha_{u, n, i}[\ell]\ k_{u, n, i, i}[\ell] + 2 \sum_{i=1}^{\mt} \sum_{j=i+1}^{\mt} \left\| k_{u, n, i, j}[\ell] \right\|  \\ &\small \times \sqrt{\alpha_{u, n, i}[\ell] \alpha_{u, n, j}[\ell]} \\
    &\small \times \cos \left( \beta_{u, n, j}[\ell] - \beta_{u, n, i}[\ell] + \arg \left\{ k_{u, n, i, j}[\ell] \right\} \right),
\end{aligned}
\label{gain1}
\end{equation}
where $\mathbf{K}_{u,n}[\ell] = \mathbf{H}_{u,n}^\daggerr[\ell] \ \mathbf{H}_{u,n}[\ell]$.
In \eqref{gain1}, $k_{u,n,i,j}[\ell]$ denotes the element in the $i^{\text{th}}$ row $j^{\text{th}}$ column of $\mathbf{K}_{u, n}[\ell]$. Meanwhile, the vector $\mathbf{f}_{u,n}$ can be expressed as 
\begin{equation} \small
\small
\begin{aligned}
   \mathbf{f}_{u, n}( \boldsymbol{\alpha}_u[\ell],\boldsymbol{\beta}_u[\ell])= \left[\alpha_{u, n, 1}[\ell] \ e^{j\beta_{u, n, 1}[\ell]},\right.\\  \left.\ldots, \alpha_{u, n, \mt}[\ell]\ e^{j\beta_{u, n, \mt}[\ell]}\right]^\top \in \mathbb{C}^{\mt \times 1}
    \label{beamvector}
    \end{aligned}
\end{equation}
where $\boldsymbol{\alpha}_u[\ell]$ denotes the power share between antennas and $\boldsymbol{\beta}_u[\ell]$ represents the phase difference between them. Based on \eqref{rateequn}, the achievable rate of user $u$ over all assigned subcarriers can be expressed as follows
\begin{equation} \small
r_{u}(\pvl) =\sum_{n=1}^{\nc} r_{u,n}(\pv[\ell]).
\label{eq:rate}
\end{equation}
\subsection{Exposure Model}
In practical uplink wireless communications, the \ac{SAR} is generally used as a metric to evaluate the exposure, which is defined as the power absorbed by the human tissue per unit mass \cite{ICNIRP}. Hence, \ac{EMF}-efficient communication networks are those endowed with low \ac{SAR} while having a satisfactory \ac{QoS}.
 
For a single-antenna \ac{UE}, the \ac{SAR} is directly proportional to the transmit power such that
\begin{equation} \small
\mathrm{SAR}_{u}[\ell]=p_{u}[\ell]\: \overline{\mathrm{SAR}} ~[\text{W/Kg
}],
\end{equation}
where $p_{u}$ is the transmit power of user $u$, and $\overline{\mathrm{SAR}}~ [\text{1/Kg}]$ is the induced reference \ac{SAR} when the transmit power is unity. The reference \ac{SAR} depends on the antenna type of \ac{UE}, the frequency of operation, the user's posture, etc. \cite{1296850}.

However, in the case of multi-antenna \ac{UE}, the relation between the \ac{SAR} and transmit power is not straightforward due to the coupling between the antennas and the user's body. Therefore, the relative share of power and the consideration of the phase shift between antennas can have a significant effect on the \ac{SAR} level \cite{heliot2020exposure}. Notably, when examining a user device with $\mt$ planar inverted F antennas (PIFA), the \ac{SAR} is expressed as a function of the transmit beamforming parameters as follows 
\begin{equation} \small
\mathrm{SAR}_{u,n}[\ell] =p_{u,n}[\ell] \: \overline{\mathrm{SAR}}(\boldsymbol{\alpha}_{u,n}[\ell],\boldsymbol{\beta}_{u,n}[\ell]),
\end{equation}
where $\boldsymbol{\alpha}_{u,n}[\ell] \in \mathbb{R}_{+}^{1\times \mt}$ and $\boldsymbol{\beta}_{u,n}[\ell] \in \mathbb{R}^{1\times \mt}$ are vectors containing the relative power
share and phase difference between the antenna ports of the $u^{\text{th}}$ \ac{UE}, respectively \cite{heliot2020exposure}. More precisely, the vector $\boldsymbol{\alpha}_{u,n}$ is defined as $\boldsymbol{\alpha}_{u,n}[\ell] \triangleq \left[\alpha_{u,n,1}[\ell], \dots, \alpha_{u,n,\mt}[\ell]\right]$
where $\alpha_{u,n,i}[\ell]\triangleq {\sqrt{p_{u,n,i}[\ell]}}/{\sqrt{p_{u,n,1}[\ell]}}, \forall i \in \left\{1,\dots,\mt\right\}$, represents the relative share of transmit power of the $i^{\text{th}}$ antenna with respect to that of the first antenna.  {In this paper, we consider a worst-case scenario based on talk position where the \ac{UE} is positioned 5 mm away from a phantom head model with realistic heterogeneous dielectric properties \cite{heliot2020exposure}}. 

For a \ac{UE} equipped with two PIFA antennas, i.e., $M_t = 2$, the induced reference \ac{SAR} is given by 
\begin{equation} \small
\small
\begin{aligned}
&\overline{\mathrm{SAR}}\left(\boldsymbol{\alpha}_{u,n}[\ell], \boldsymbol{\beta}_{u,n}[\ell]\right) 
= b_{1} \alpha_{u,n,1}[\ell] 
+ b_{2} \sqrt{\alpha_{u,n,1}[\ell] \alpha_{u,n,2}[\ell]} 
\\
&+ b_{3} \alpha_{u,n,2}[\ell]  + \bigg(b_{4} \alpha_{u,n,1}[\ell] 
+ b_{5} \sqrt{\alpha_{u,n,1}[\ell] \alpha_{u,n,2}[\ell]} 
\\&\quad + b_{6} \alpha_{u,n,2}[\ell] \bigg) 
 \times \sum_{i=7}^{13} b_{i} \cos \big((i-7) \beta_{u,n,2}[\ell] + b_{i+7}\big),
\end{aligned}
\label{eq:SARm}
\end{equation}
where the parameters in $\mathbf{b}=[b_1,b_2,\cdots, b_{20}]$ are chosen to fit SAR values derived from the simulation of SAR distribution within the phantom head \cite{heliot2020exposure}.
Hence, we can see from \eqref{eq:SARm} that the \ac{SAR} depends directly on the relative power and the phase shift of the signal passed to different antennas. Typically, these values are designed to shape the antenna pattern and beamform the signal towards a specific direction. Therefore, the beamforming angle will have an indirect impact on the exposure. To illustrate this phenomenon, we show in Fig.~\ref{fig:exp/var} the induced \ac{SAR} as a function of the direction of the main lobe of the antenna array. We can see that designing $\boldsymbol{\alpha}$ and $\boldsymbol{\beta}$ to achieve a specific direction of the main antenna array lobe can have a significant impact on the induced \ac{SAR}, emphasizing the necessity of considering these parameters in the design of \ac{EMF}-aware uplink \ac{MIMO} systems. 
For a multi-carrier system, we quantify the individual exposure of user $u$ by the total \ac{SAR} over all subcarriers as follows
\begin{equation} \small
\small
E_{u}\left(\pvl \right)=\sum_{n=1}^{\nc} \delta_{u, n}[\ell] \ p_{u,n}[\ell]~ \overline{\mathrm{SAR}} \left(\boldsymbol{\alpha}_{u, n}[\ell], \boldsymbol{\beta}_{u, n}[\ell]\right),
\label{EIt}
\end{equation}
where $\pvl$ denote the system parameters at $\ell$ time. Based on the per-user exposure at a specific time slot,  {we define the average uplink exposure across all the users as the mean value of the uplink \ac{EMF} exposure levels experienced by individual users in the network over time, as follows} \cite{TethredUAVAhmed}
\begin{equation} \small
\small \EI \left(\Gamma \right) \triangleq \frac{\Delta}{N_T U} \sum_{\ell=1}^{N_T} \sum_{u=1}^{U} E_{u}\left(\boldsymbol{\Gamma}[\ell] \right)~[\text{W/Kg}].
\label{TotalExp}
\end{equation}
 As shown in \eqref{TotalExp}, the exposure is presented as a function of all the system variables which can be efficiently exploited to have an \ac{EMF}-aware design. The dependence on these parameters provides various degrees of freedom that can be leveraged to reduce the exposure level.

In summary, the \ac{EMF} exposure, especially for multi-antenna \ac{UE}, relies on both \ac{UE} transmit power and reference \ac{SAR} value which depends on the antenna configuration. The interplay of $\boldsymbol{\alpha}$ and $\boldsymbol{\beta}$ along with transmit power highlights the importance of incorporating these factors into \ac{EMF}-aware \ac{MIMO} system design, promoting safer and more efficient communication networks.
\subsection{Problem Formulation}
We investigate a strategy design for the \ac{ARIS}-assisted multi-user \ac{MIMO} uplink transmission with reduced \ac{EMF} exposure, where we aim to optimize the parameters ${\pv}$ while maintaining the \ac{QoS} for all the network users in terms of the required uplink data rate. Considering the relevant constraints, our optimization problem can be formulated as follows
\begin{subequations} \small
\begin{alignat}{2}
&(\textrm{P2}) \quad \underset{\pv}{\textrm{minimize}} \quad \EI \left(\pv\right), \label{eq:optProb}\\
&\quad \textrm{subject to:} \nonumber \\
&\quad r_{u} \left(\pvl\right) \geq \bar{R}_{u}, \quad \forall u \in \mathcal{U}, \forall \ell \in \mathcal{N_T}, 
\label{eq:QoSConstraint}\\
&\quad \sum_{u=1}^{U} \delta_{u,n}[\ell] \leq 1, \quad \forall n \in \mathcal{N}_c, \forall \ell \in \mathcal{N_T}, 
\label{REConstraint}\\
&\quad |\theta_i[\ell]| = 1, \quad \forall i \in \mathcal{N}, \forall \ell \in \mathcal{N_T}, 
\label{RISConstraint}\\
&\quad \sum_{n=1}^{\nc} \delta_{u,n}[\ell] \, p_{u,n}[\ell] \leq P_{\text{max}}, \quad \forall u \in \mathcal{U}, \forall \ell \in \mathcal{N_T}, 
\label{eq:PowerConstraint}\\
&\quad \|\mathbf{q}[\ell+1] - \mathbf{q}[\ell]\|^2 \leq D_{\max}^2, \quad 1 \leq \ell \leq N_T - 1, 
\label{UAV-dis}\\
&\quad \mathbf{q}[1] = \mathbf{q}_0, \quad \mathbf{q}[N_T] = \mathbf{q}_T. 
\label{finalpoint}
\end{alignat}
\end{subequations}
The constraint \eqref{eq:QoSConstraint} ensures that each user $u$ satisfies his required data rate $\bar{R}_{u}$ and the constraint \eqref{eq:PowerConstraint} assures that the transmit power of each user $u$ is below the maximum supported transmit power $P_{\text{max}}$. Meanwhile, the \ac{RE} allocation constraint presented in \eqref{REConstraint} ensures that any \ac{RE} can only be allocated to one user at a time.
Furthermore, the constraints \eqref{UAV-dis} regulate the distance between two consecutive locations on the trajectory to be below the maximum distance that the \ac{ARIS} can fly within a time slot $D_{\max}$. Further, the constraint \eqref{finalpoint} specifies the initial and final locations for the \ac{UAV} as $\mathbf{q}_o$ and $\mathbf{q}_T$, respectively.

Note that it is difficult to directly tackle (P1) due to the introduction of the multi-antenna \ac{SAR} model. More specifically, in our scheme, the optimization of the \ac{EMF} exposure can not be simplified into a traditional power allocation problem. Additionally, the non-convex and discrete constraints, \eqref{eq:QoSConstraint}, \eqref{REConstraint}, and \eqref{RISConstraint}, further complicate the joint optimization. 
\section{Proposed Algorithm for EMF-Efficient MU-MIMO Network with ARIS}
\label{sec:PbFormulation}
Due to the coupling of the optimization variables, it is complicated to optimize the set of variables $\pv$ jointly, especially with high dimensional parameters and non-convex constraints. To reduce complexity, we apply the \ac{AO} method, which is an iterative approach optimizing one variable while fixing the other variables alternately.
\subsection{MIMO Parameters Optimization}
In this subsection, we optimize the \ac{MIMO} parameters $\boldsymbol{\alpha}$ and $\boldsymbol{\beta}$. It is worth noting that the beamforming vectors designed at a specific time instant are independent of those designed at different time slots and from those of other users. 
Therefore, we solve this subproblem by optimizing the individual exposure at each time slot for each user separately.

Suppose the \ac{RE} allocation vector $\boldsymbol{\delta}$, phase-shift matrix $\boldsymbol{\Theta}$ and \ac{ARIS} location $\mathbf{q}$ are fixed in the feasible set that satisfies the corresponding constraints. 
At time $\ell$, the problem minimizing the exposure \ac{w.r.t}  relative power share $\boldsymbol{\alpha}_u$ and antenna shift $\boldsymbol{\beta}_u$ between the antennas ports of \ac{UE} $u$ is formulated as follows 
\begin{subequations} \small
\begin{alignat}{3}
(\textrm{P2}) \quad &\underset{\boldsymbol{\alpha}_u,\boldsymbol{\beta}_u }{\textrm{minimize}}        &\quad& E_u\left(\pvl\right), & \forall \ell \in \mathcal{N_T}, \forall u \in \mathcal{U},\\
&\textrm{subject to} &                & \eqref{eq:QoSConstraint} \label{rateconstraint}.
\end{alignat}
\end{subequations}
However, to initialize our \ac{AO} iterative algorithm, we assume that all the users achieve their minimum required rate. By making this assumption, we ensure that the \ac{QoS} constraint \eqref{eq:QoSConstraint} is satisfied. Additionally, we assume an equal share of data rate among subcarriers, which translates to an equal power distribution among the allocated subcarriers, which is not necessarily the optimal power allocation. These initial assumptions ensure the feasibility of the problem. Subsequently, the optimal distribution of required
rates across subcarriers is determined by computing the optimal power allocation for each
subcarrier to achieve an \ac{EMF}-efficient power control mechanism in Sec.~\ref{sec:transmitpoweropt}. More in detail, the minimum transmit power that satisfies the rate constraint can be written from \eqref{rateequn} and \eqref{eq:QoSConstraint} as 
\begin{equation} \small
    p_{u,n}(\pvun) = (2^{\bar{r}_{u,n}/w}-1) \frac{\sigma^2}{\gamma (\pvun)},
    \label{power1}
\end{equation} 
where $\bar{r}_{u,n}$ is the minimum required data rate through the $n^\text{th}$ subcarrier. For notation convenience, we remove the dependency on time from the notation. 
Substituting \eqref{power1} into the expression of the exposure, we get
\begin{equation} \small
\begin{split}
 E_{u}(\pv) =\sum_{n=1}^{\nc} \delta_{u, n} (2^{\bar{r}_{u,n}/w}-1)  \frac{\sigma^{2} \operatorname{\overline{\mathrm{SAR}}}\left(\boldsymbol{\alpha}_{u, n}, \boldsymbol{\beta}_{u, n}\right)}{ \gamma \left(\pvun\right)}.
\label{expbeam}
\end{split}
\end{equation}
For a predefined $\bar{r}_{u,n}$, the beamforming parameters over different subchannels are independent of each other. Hence, the problem is equivalent to separately minimizing the ratio presented in \eqref{expbeam} for each \ac{RE} $n$ as follows
\begin{subequations} \small
\begin{alignat}{2}
\underset{\boldsymbol{\alpha}_{u,n},~\boldsymbol{\beta}_{u,n}}{\textrm{minimize}} 
&\quad \delta_{u, n} \sigma^{2} (2^{\bar{r}_{u,n}/w}-1)  \frac{\operatorname{\overline{\mathrm{SAR}}}\left(\boldsymbol{\alpha}_{u, n}, \boldsymbol{\beta}_{u, n}\right)}{ \gamma \left(\pvun\right)} \\
\textrm{subject to}&\quad \boldsymbol{\alpha}_{u, n} \in \mathbb{R}_{+}^{\mt \times 1},\\  
&\quad \boldsymbol{\beta}_{u, n} \in \mathbb{R}^{\mt \times 1}.
\end{alignat}
\end{subequations}
The formulated problem is a non-convex single-ratio programming. Therefore, we consider the Dinkelbach technique to decouple the denominator and nominator \cite{fractionalProg}. This method converts the ratio into a linear parameterized expression, which is then optimized iteratively, where the slack parameter $\lambda$ is updated at each iteration until convergence. At each iteration $i$, we solve the following problem
\begin{subequations} \small
\begin{alignat}{2}
\underset{\boldsymbol{\alpha}_{u,n},~\boldsymbol{\beta}_{u,n}}{\textrm{minimize}} 
&\quad f_1\left(\boldsymbol{\alpha}_{u, n},\boldsymbol{\beta}_{u, n},\lambda^{(i)}\right) \\
\textrm{subject to}&\quad \boldsymbol{\alpha}_{u, n} \in \mathbb{R}_{+}^{\mt \times 1},\\  
&\quad \boldsymbol{\beta}_{u, n} \in \mathbb{R}^{\mt \times 1},
\end{alignat}
\label{MIMOOptPb}
\end{subequations}
where the objective function depends on the slack variable $\lambda^{(i)}$ as follows
\begin{equation} \small
\begin{aligned}
f_1\left(\boldsymbol{\alpha}_{u, n},\boldsymbol{\beta}_{u, n},\lambda^{(i)}\right)= \delta_{u,n}\  \left(\sigma^{2} \left(2^{\bar{r}_{u,n}/w}-1\right) \right.\\ \\ \left.\operatorname{\overline{\mathrm{SAR}}}\left(\boldsymbol{\alpha}_{u, n}, \boldsymbol{\beta}_{u, n}\right)- \lambda^{(i)} \gamma \left(\pvun\right)\right),
\end{aligned}
\end{equation}
and $\lambda^{(i)}$ is updated as 
\begin{equation} \small  \lambda^{(i)}= \delta_{u,n}~
\sigma^{2} (2^{\bar{r}_{u,n}/w}-1)\frac{\operatorname{\overline{\mathrm{SAR}}}\left(\boldsymbol{\alpha}^{\ast (i-1)}_{u, n}, \boldsymbol{\beta}^{\ast (i-1)}_{u, n}\right)}{ \gamma \left(\pvun^{\ast(i-1)}\right)}.
\label{lambdaMIMO}
\end{equation}
The solution of this subproblem is detailed in the iterative Algorithm~\ref{MIMOAlgo}, which guarantees the convergence to an optimal solution. The complexity of the proposed algorithm is with the order of $\mathcal{O}\left( \sqrt{2\mt-2}\right)$.
\begin{algorithm}[t!]
\caption{EMF Exposure-Aware Beamforming Design}
\begin{algorithmic}[1]
\State Initialize $\lambda^{(0)}$, iteration index $i=0$, convergence threshold $\epsilon_1 \geq 0$.
\State \textbf{repeat}
\State Calculate $\left(\boldsymbol{\alpha^\ast}_{u,n}, \boldsymbol{\beta^\ast}_{u,n}\right)^{(i)}$ by solving \eqref{MIMOOptPb}.
\State Update $\lambda^{(i+1)}$ according to Eq.~\eqref{lambdaMIMO}.
\State Set $i = i + 1$.
\State \textbf{until} $\left| \lambda^{(i)} - \lambda^{(i-1)}\right| \leq \epsilon_1$.
\end{algorithmic} 
\label{MIMOAlgo}
\end{algorithm}
\subsection{RIS Phase-shift Vector Design}
Consider the optimization variables being fixed, the optimization of (P1) \ac{w.r.t} the \ac{RIS} vector $\boldsymbol{\theta}$ is formulated as
\begin{subequations} \small
\begin{alignat}{2}
(\textrm{P3})~& \underset{\boldsymbol{\theta} ~\in ~\mathbb{C}^{N\times 1}}{\textrm{minimize}} &~&  \sum_{u=1}^{U} \sum_{n=1}^{\nc} \delta_{u,n}\frac{  \sigma^2  \left(2^{\bar{r}_{u,n}/w}-1\right)\operatorname{\overline{\mathrm{SAR}}}_{u,n}}{\gamma (\pvun)} \\
    & \text{subject to}&      &  \eqref{RISConstraint}. \end{alignat}
\end{subequations}
It is worth noting that the phase shifts of the ARIS are designed only to assist the uplink transmission for users within the considered cell.
To simplify the notation, let $\operatorname{\overline{\mathrm{SAR}}}_{u,n}$ denote $\operatorname{\overline{\mathrm{SAR}}}\left(\boldsymbol{\alpha}_{u,n}, \boldsymbol{\beta}_{u, n}\right)$.
{Due to the non-convex nature of the problem, which arises from its formulation as a sum-of-ratios problem, we employ the quadratic transform technique to reformulate it \cite{fractionalProg}. This technique aims to recast the non-convex sum of fractions problem as a sequence of convex problems that can be efficiently solved. As such, the problem can be reformulated as}
\begin{subequations} \small
\begin{alignat}{2}
(\textrm{P3}) \quad & \underset{\boldsymbol{\theta},\mathbf{y}}{\textrm{minimize}} &\quad&  f_2 \left(\boldsymbol{\theta},\mathbf{y}\right) 
    \label{f}\\
    & \text{subject to}&      &  \eqref{RISConstraint},  \end{alignat}
\end{subequations}
where the new objective function is expressed as
\begin{equation} \small
    f_2 \left(\boldsymbol{\theta},\mathbf{y}\right)\triangleq \sum^U_{u=1}\sum^{\nc}_{n=1}  \delta_{u, n} \left(2~ y_{u,n}~  c_{u,n} - y_{u,n}^2 \gamma\left(\pvun\right)\right),
\end{equation} 
and
\begin{equation} \small
    c_{u,n} \triangleq \sqrt{\sigma^2 \, (2^{\bar{r}_{u,n}/w}-1)~ \overline{\mathrm{SAR}}_{u,n}}.
\end{equation} Here, $\mathbf{y}=[y_{1,1},..,y_{u,n},..,y_{U,N}]$ refers to the vector of auxiliary variables.
Resorting to the \ac{AO} method, we address the new formulated problem by iteratively optimizing the introduced variables in $\mathbf{y}$. The optimal value for $\mathbf{y}$ is obtained by solving $\partial{f_2(\boldsymbol{\theta},\mathbf{y})}/\partial{\mathbf{y}}=0$, while fixing $\boldsymbol{\theta}$ as follows
\begin{equation} \small
y_{u,n}^\ast\triangleq\frac{{c_{u,n}}}{\gamma(\pvun)}.
\label{YPhase}
\end{equation}
Subsequently, the objective function can be written as 
\begin{equation} \small
  f_2 \left(\boldsymbol{\theta},\mathbf{y}\right) = \sum^U_{u=1}\sum^{\nc}_{n=1}  \delta_{u,n} m_{u,n}-\delta_{u,n} y_{u,n}^2 \|\mathbf{H}_{n} \ \boldsymbol{\Theta} \mathbf{g}_{u,n}+\mathbf{h}^{\text{d}}_{u,n}\|^2,
  \label{f4}
\end{equation}
where $m_{u,n} = 2 y_{u,n} c_{u,n}$, $\mathbf{g}_{u,n}\triangleq \mathbf{G}_{u,n} \ \mathbf{f}_{u,n}$ and $\mathbf{h}^{\text{d}}_{u,n} \triangleq \mathbf{H}^{\text{d}}_{u,n}\mathbf{f}_{u,n}$. To solve this problem optimizing \ac{RIS} phase shifts while fixing $\mathbf{y}$, we use the matrix lifting technique \cite{semidefiniteRelaxation}. First, we rewrite the objective function differently as follows
\begin{equation} \small
 \begin{aligned}
 \small
&f_2(\boldsymbol{\theta},\mathbf{y})= \sum^U_{u=1}\sum^{\nc}_{n=1} \delta_{u,n}  m_{u,n} \\&- \delta_{u,n} y_{u,n}^2 \left( \boldsymbol{\theta}^\daggerr \mathbf{A}_{u,n} \boldsymbol{\theta} + 2\mathrm{Re}\left\{\boldsymbol{\theta}^\daggerr \mathbf{b}_{u,n}\right\}  \right),
\end{aligned}
\end{equation}
where $\mathbf{A}_{u,n} \triangleq \mathrm{diag}\left(\mathbf{g}_{u,n}\right)^\daggerr~ \mathbf{H}_{n}^\daggerr~  \mathbf{H}_{n} ~\mathrm{diag}\left(\mathbf{g}_{u,n}\right),$ and $\mathbf{b}_{u,n} \triangleq \mathrm{diag}\left(\mathbf{g}_{u,n}\right)^\daggerr~ \mathbf{H}_{n}^\daggerr ~\mathbf{h}^{\text{d}}_{u,n}$.
By removing all the terms that do not depend on $\boldsymbol{\theta}$, we can write the optimization problem for fixed $\mathbf{y}$ as 
\begin{subequations} \small
\label{32}
\begin{alignat}{2}
&\underset{{\boldsymbol{\theta}}}{\textrm{minimize}}        &\quad& -{\boldsymbol{\theta}}^\daggerr \boldsymbol{R} ~\boldsymbol{\theta} \\
&\textrm{subject to} &      & |\boldsymbol{\theta}_i| = 1, \forall  i=\{1,..,N+1\}.
\end{alignat}
\end{subequations}
Here, the matrix $\mathbf{R}$ is defined as $\mathbf{R} \triangleq \begin{bmatrix}
    \mathbf{A} & \mathbf{b} \\
    \mathbf{b}^\daggerr & 0 
    \end{bmatrix},$ where 
    \begin{equation}
\small
\begin{aligned}
    \mathbf{A} &\triangleq \sum_{u=1}^{U} \sum_{n=1}^{\nc} \delta_{u,n} \, y_{u,n}^2 \, \mathbf{A}_{u,n}, \\
    \mathbf{b} &\triangleq \sum_{u=1}^{U} \sum_{n=1}^{\nc} \delta_{u,n} \, y_{u,n}^2 \, \mathbf{b}_{u,n}.
\end{aligned}
\end{equation}
The reformulated problem is non-convex \ac{QCQP} due to the unit modulus constraint. Thus, we proceed by using the matrix lifting technique to reformulate \eqref{32} as a rank-one optimization problem and we rewrite the problem \ac{w.r.t} the matrix $\boldsymbol{\bar{\Theta}}=\bar{\boldsymbol{\theta}}\bar{\boldsymbol{\theta}}^\daggerr$, where $\bar{\boldsymbol{\theta}}=\begin{bmatrix}
\boldsymbol{\theta}\\
1
\end{bmatrix}$. Let us rewrite \eqref{32} in terms of $\bar{\boldsymbol{\Theta}}$
as
\begin{subequations} \small
\begin{alignat}{2}
(\textrm{P3}^\prime) \quad &\underset{\mathbf{\bar{\Theta}}}{\textrm{minimize}}\quad &\quad& - \mathrm{tr}\left( \mathbf{R} \boldsymbol{\bar{\Theta}}\right) \\
&\textrm{subject to} &\quad& \boldsymbol{\bar{\Theta}}_{i,i} = 1, \forall i=\{1,..,N+1\},\label{UnityPhaseConst}\\
& \quad & \quad& \boldsymbol{\bar{\Theta}} \succeq  0.\label{PSPhaseConst}
\end{alignat}
\end{subequations}
\begin{algorithm}[t!]
\caption{EMF Exposure-Aware RIS Phase Shifts Design}
\label{alg:PhaseShift}
\begin{algorithmic}[1]
\State Initialize $\mathbf{y}^{(0)}$, set iteration index $i=0$, \text{ number of} Gaussian randomization $I_{\text{GR}}$,and convergence threshold $\epsilon_2 \geq 0$.
\State \textbf{repeat}
  \State \quad Solve problem $(\textrm{P3}^\prime)$ using SDR for fixed $\mathbf{y}^{(I_{2})}$, $\boldsymbol{\bar{\Theta}}^\ast$, and $I_{\text{GR}}$.
  \State \quad \textbf{for} $k = 1, \ldots, I_{\text{GR}}$  
    \State \qquad Generate a matrix with zero mean and $\boldsymbol{\bar{\Theta}}^\ast$ variance: $\boldsymbol{\eta}^k \doteq \mathcal{N}(0, \boldsymbol{\bar{\Theta}}^\ast)$.
    \State \qquad Construct the feasible vector as $(\boldsymbol{\theta}^k)^\ast = \text{sgn}(\boldsymbol{\eta}^k)$.
  \State \quad \textbf{End}
  \State \quad Determine the approximate solution of $\boldsymbol{\theta}^\ast = \underset{\hat{\boldsymbol{\theta}}_i}{\text{min }} \boldsymbol{\theta}_i^\top \mathbf{R} \boldsymbol{\theta}_{i}$.
  \State \quad Set $i = i + 1$.
  \State \quad Update $\boldsymbol{y}^{(i)}$ according to Eq.~\eqref{YPhase}.
\State \textbf{until} $|\boldsymbol{y}^{(i-1)} - \boldsymbol{y}^{(i)}| \leq \epsilon_2$.
\end{algorithmic}
\end{algorithm}
 Considering that $\boldsymbol{\bar{\Theta}}$ is a positive semi-definite matrix with each diagonal element equal to unity, we introduce the constraint \eqref{UnityPhaseConst} and \eqref{PSPhaseConst}.
Now the problem is formulated as a standard convex \ac{SDP} allowing us to relax the rank-one constraint and optimally solve it by an existing convex optimization solver such as CVX \cite{CVX}.
In general, a relaxed problem may not lead to a rank-one solution, which means that the obtained solution serves only as an upper bound of the optimal solution of the problem $(\textrm{P3}^\prime)$. Therefore, we use Gaussian randomization to construct a rank one matrix to the initial problem $(\textrm{P3})$ from the optimal solution of $(\textrm{P3}^\prime)$.
Initially, we derive the eigenvalue decomposition of $\boldsymbol{\Theta}=\mathbf{U}\mathbf{D}\mathbf{U}^\daggerr$ where $U$ is a unitary matrix and $\mathbf{D} = \mathrm{diag}\left(\lambda_1 \dots \lambda_{N+1}\right)$ is a diagonal matrix, both with same size $(N+1)\times(N+1)$. 
Then, we have a first suboptimal solution for $(\textrm{P3}^\prime)$ as $\boldsymbol{\bar{\boldsymbol{\theta}}}=\mathbf{U}D^{\frac{1}{2}}\mathbf{r}$, where $\mathbf{r} \in \mathbb{C}^{(N+1)\times 1}$ is a random vector generated according to a circularly symmetric complex Gaussian (CSCG) distribution with zero-mean and $\boldsymbol{I}_{N+1}$ covariance matrix.
Furthermore, we approximate the optimal solution of $(\textrm{P3})$ by generating independently random Gaussian vectors $\mathbf{r}$.
Finally, we recover the solution $\boldsymbol{v}$ based on 
 $\boldsymbol{\theta} = e^{j \arg \left(\frac{\boldsymbol{\bar{\boldsymbol{\theta}}}}{\boldsymbol{\bar{\boldsymbol{\theta}}}_{N+1}}\right)}.$
It has been proved that following the \ac{SDR} approach by a sufficiently large number of randomization of $\mathbf{r}$ guarantees an $\frac{\pi}{4}$-approximation of the optimal objective function value of the initial problem \cite{semidefiniteRelaxation}. 
Regarding the complexity of the proposed solution, it is worth noting that SDP is typically solved using the interior point method which yields an $\epsilon$-optimal solution. Hence, by reserving the term with highest order and assuming perfect \ac{CSI}, the complexity of the proposed solution is $\mathcal{O}\left( N^{3.5}\right)$.\footnote{The channel estimation process in \ac{RIS}-aided networks can be obtained based on existing methods, such as the techniques outlined in \cite{RISChannelEst,ChannelEstRIS,9956764}.}
\subsection{RE Allocation Optimization}
Finding the optimal number of allocated subcarriers to each user is essential to reduce the overall exposure effectively. It is worth noting that the allocation of multiple subcarriers to a single user has a remarkable effect of diminishing the necessary transmission power, and consequently reducing the individual exposure \eqref{EIt}. In this regard, for given \ac{RIS} phase shifts, beamforming parameters, and \ac{ARIS} trajectory, the resource allocation problem can be formulated as 
\begin{subequations} \small
\begin{alignat}{2}
(\textrm{P4}) \quad & \underset{\boldsymbol{\delta}\in \mathbb{R}^{U\times \nc}}{\textrm{minimize}} && \quad \EI (\boldsymbol{\pv}) \\
& \text{subject to} && \quad  \eqref{eq:PowerConstraint},\eqref{REConstraint}.
\end{alignat}
\label{eq.REA}
\end{subequations}
This subproblem is a constrained binary non-convex problem which is challenging to solve. Therefore, we propose a heuristic algorithm for resource allocation. Initially, each user is allocated one \ac{RE}, then, further subcarriers are assigned to the users with the highest expected \ac{EMF} exposure, quantified by a proper metric. We establish a ranking metric designed to exhibit scaling laws analogous to those of the \ac{EMF} exposure with respect to the required data rate and path-loss, i.e., 
 \begin{equation} \small
\rho_{u}^{(i)} \triangleq (2^{r_u^{(i)}/w}-1)(d_{u\text{R}}^{\kappa_1}~ d_{\text{RB}}^{\kappa_2}),
\end{equation}
where $r_u^{(i)} = r_u^\ast /\mathrm{N}_u^{(i)}$ quantifying the rate achieved through the allocated subcarriers $\mathrm{N}_u^{(i)}$ which is updated at each iteration $i$ of the algorithm. However, the total rate $r_u^\ast$ is achieved through prior subproblems optimization which ensures that power remains within the maximum allowed budget constraint. Given the optimized variables alternately updated, we ensure that the power does not exceed the maximum allowed power limit.
This metric is proportional to the \ac{EMF} exposure which prioritizes users that have a high required rate and path-loss. At the $i^{\text{th}}$ iteration, we assign a subcarrier to the user with the highest value of $\rho_{u}^{(i)}$. Then, we update the value of the metric and repeat the allocation process till all the remaining subcarriers are assigned to users, as detailed in Algorithm \ref{alg:RE}. The complexity of this algorithm is in the order of $\mathcal{O}\left(\nc-U\right)$.
\begin{algorithm}[t!]
\caption{EMF Exposure-Aware Resource Element Allocation}
\label{alg:RE}
\begin{algorithmic}[1]
\State Initialize iteration index $i = 1$. 
\State Assign one \ac{RE} to each user, i.e., $\mathcal{N}_u = \{u\}, \forall u \in \mathrm{U}$.
\State Define $\mathcal{S}$ as the set of non-assigned subcarriers: $\mathcal{S} = \left\{U+1, \dots, \nc\right\}$.
\State \textbf{while} $\mathcal{S} \neq \emptyset$   
\State \quad Solve $u^\ast = \underset{u \in \mathrm{U}}{\text{argmax}} \left(\rho_{u}^{(i)}\right)$.
\State \quad Assign the first subcarrier from $\mathcal{S}$ to user $u^\ast$, i.e., $\mathcal{N}_{u^\ast} = \mathcal{N}_{u^\ast} \cup \: \mathcal{S}(1)$.
\State \quad Update $\delta_{u^\ast,\mathcal{S}(1)} = 1$.
\State \quad Remove the assigned subcarrier from $\mathrm{S}$,
$ \text{i.e., } \mathcal{S} = \mathcal{S} \setminus \mathcal{S}(1)$.
\State \textbf{end}
\end{algorithmic}
\end{algorithm}
\subsection{Power control}
\label{sec:transmitpoweropt}
As we mentioned before, we initially assumed an equal distribution of transmit power among subcarriers to ensure feasibility and satisfy QoS constraints. However, this allocation may not be optimal.
Therefore, a proper design for the power sharing between subcarriers is essential to minimize the exposure. Once we obtain the optimal power allocation, we update the data rate achieved at each subcarrier. Then, we solve other subproblem iteratively until convergence is reached. In the following, we consider the optimization of (P1) \ac{w.r.t} the transmit power $\mathbf{P}$ while holding the remaining optimization variables fixed.
Given the independence of each user's transmit power, wherein each user can adjust their power level without influencing or being influenced by the power levels of other users, we proceed to optimize the power allocated to each user individually for different time slots as follows
\begin{subequations} \small
\begin{alignat}{3}
 (\textrm{P5}) \quad & \underset{\mathbf{p}_u \in \mathbb{R}_+^{1\times \nc}} {\textrm{minimize}}        &\quad& \sum_{n=1}^{\nc} \delta_{u,n} \  \overline{\mathrm{SAR}}_{u,n} \ p_{u,n}\\
&\textrm{subject to} &      &r_{u} \left(\pv\right) \geq   \bar{R}_{u}, \, \label{QoSConstraint}\\
&                  &      & \sum_{n=1}^{\nc} \delta_{u, n}\ p_{u,n} \leq P_{\text{max}}.\label{PowerConstraint}
\end{alignat}
\label{powPb}
\end{subequations}
The subproblem (\textrm{P5}) is a convex nonlinear optimization problem \ac{w.r.t} $\mathbf{p}_{u}$ which can be solved using the KKT conditions. The expression of the optimal transmit power over the $n^{\text{th}}$ subcarrier can be derived as
\begin{equation} \small
\begin{aligned}
\small 
p_{u,n}^\ast = \max\left\{\frac{ \delta_{u,n} w \mu^\ast}{\ln(2)(\overline{\mathrm{SAR}}_{u,n}+\lambda^\ast)}-\frac{\delta_{u,n} \sigma^2}{\gamma(\pv^\ast_{u,n})},0\right\},
\\ \small \forall u \in \mathcal{U}, \forall n \in \mathcal{N}_c,
    \label{optPower}
    \end{aligned}
\end{equation}
where $\mu^\ast$ and $\lambda^\ast$ are the Lagrangian multipliers obtained by solving the KKT equations. The details of the derivation can be found in Appendix A. The optimal power, $p_{u,n}^\ast$, allocated to each subcarrier determines the share of this subcarrier from the total required rate, i.e., $\bar{r}_{u,n}$ using \eqref{rateequn}, which is used for later iterations of the alternate optimization. Algorithm~\ref{alg:Power} outlines the proposed solution for power allocation over subcarriers. 

The complexity of the transmit power allocation is primarily associated with the computational expense of calculating the Lagrangian multipliers. In our subproblem, the complexity of the proposed solution is based on solving a system of $(\mathrm{N}_u+4)$ nonlinear equations. Hence, the complexity of Algorithm~\ref{alg:Power}, optimizing the power for all users, is in the order of $\mathcal{O}\left(\displaystyle\sum_{u=1}^{U} \mathrm{N}_u^2\right)$.
\begin{algorithm}[t!]
\caption{EMF Exposure-Aware Power Allocation}
\label{alg:Power}
\begin{algorithmic}[1]
\State Initialize $\mu^{(0)}$, $\lambda^{(0)}$, and the convergence threshold for Newton-Raphson algorithm
\State Define $\mathcal{L}$ using Eq.~\eqref{lagrange}.
\State Calculate $\mu^\ast$ and $\lambda^\ast$ solving Eq.~\eqref{SysEqPowOpt} using Newton-Raphson algorithm.
\State \textbf{for} $n = 1, 2, \dots, \nc$  
\State \quad Calculate $p_{u,n}^\ast$ using Eq.~\eqref{optPower}.
\State \textbf{end}
\end{algorithmic}
\end{algorithm}
\subsection{ARIS Trajectory Design}
After solving the optimization variables ($\boldsymbol{\alpha},\boldsymbol{\beta}$), with $\boldsymbol{\delta}$, and $\mathbf{P}$, problem (P1) regarding the optimization target $\mathbf{Q}$ is formulated as
\begin{subequations} \small
\begin{alignat}{2}
(\textrm{P6}) \quad &\underset{\mathbf{Q}\in \mathbb{R}^{2\times N_T}}{\textrm{minimize}}        &\quad& \EI\left(\pv\right)\label{eq:optProb}\\
&\textrm{subject to} &      & \eqref{UAV-dis} , \eqref{finalpoint}.
\end{alignat}
\end{subequations}
Because of the non-convexity of the objective function \eqref{eq:optProb} and the constraint \eqref{eq:QoSConstraint},\eqref{UAV-dis}, there exist great challenges to deal with (P6) directly. It can be observed that (P6) is a sum-of-ratio fractional programming. According to \cite{fractionalProg}, (P6) is equivalent to a minimization problem formulated by 
\begin{subequations} \small
\label{35}
\begin{alignat}{3}
 \quad &\underset{\mathbf{Q},\mathbf{X}}{\textrm{minimize}}         &\quad& f_4 (\mathbf{Q},\mathbf{X})\\
   &\text{subject to} &      &  \eqref{UAV-dis} , \eqref{finalpoint},
   \end{alignat}
     \end{subequations}
     where
     \begin{equation} \small  
     \begin{aligned}
     \small
     f_4 (\mathbf{Q},\mathbf{X}) \triangleq \sum^{N_T}_{\ell =1} \sum^{U}_{u=1} \sum^{\nc}_{n=1} \delta_{u,n}[\ell] 2\ x_{u,n}[\ell] \ c_{u,n}[\ell]\\ \small -\delta_{u,n}[\ell] x^2_{u,n}[\ell] \: \gamma(\pvun [\ell]).
     \end{aligned}
     \end{equation}
The matrix $\mathbf{X}[\ell] \in \mathbb{R}^{U \times \nc}, \forall \ell \in \mathcal{N_T}$, is employed as an auxiliary variable. Note that \eqref{35} is minimized by iteratively updating $\mathbf{Q}$ and $\mathbf{X}$ until convergence. For a fixed $\mathbf{Q}$, the optimal $x_{u,n}$ can be derived in closed-form as
\begin{equation} \small
\small x_{u,n}^\ast [\ell] =\frac{c_{u,n}[\ell]}{\gamma(\pv_{u,n}^\ast [\ell])} , \forall u \in \mathcal{U}, \forall n \in \mathcal{N}_c \text{ and } \forall \ell \in \mathcal{N_T}.
    \label{XTraj}
\end{equation}
However, the new equivalent problem is still non-convex with respect to $\mathbf{Q}$. To show the dependency of the objective function on the \ac{ARIS} location, we express $\gamma(\pvun [\ell])$ differently as 
\begin{equation} \small
\small    \gamma(\pvun [\ell])=\frac{a_{u,n}[\ell]}{d_{u\text{R}}^{\kappa_1}[\ell]~d_{\text{RB}}^{\kappa_2}[\ell]} + \frac{b_{u,n}[\ell]}{\sqrt{d_{u\text{R}}^{\kappa_1}[\ell]~d_{\text{RB}}^{\kappa_2}[\ell]}},
\end{equation}
where
\begin{equation}
    \small a_{u,n}[\ell]= \rho^2\ \| \bar{\mathbf{H}}_n[\ell] \boldsymbol{\Theta}[\ell] \bar{\mathbf{G}}_{u,n}[\ell]\mathbf{f}_{u,n}[\ell]\|^2,
\end{equation}
\begin{equation}
    \small b_{u,n}[\ell]= 2\ \rho \ \mathrm{Re}\left\{\mathbf{h}_{u,n}^\text{d}[\ell]^\daggerr \bar{\mathbf{H}}_n[\ell] \boldsymbol{\Theta}[\ell] \bar{\mathbf{G}}_{u,n}[\ell]\mathbf{f}_{u,n}[\ell] \right\},
\end{equation}
and
\begin{equation}
    \small \mathbf{h}_{u,n}^\text{d}[\ell] = \mathbf{H}_{u,n}^\text{d}[\ell] \mathbf{f}_{u,n}[\ell].
\end{equation}
It is noteworthy that the non-convexity of the subproblem comes from the coupling between the distances between users-\ac{ARIS} and \ac{ARIS}-\ac{BS}. With this in mind, we intend to decouple the two variables and relax the problem by introducing new auxiliary variables $\mathbf{U}$ and $\mathbf{v}$, which establish bounds on the distances. Specifically, $\mathbf{U}$ bounds the distance between users and the \ac{ARIS}, and $\mathbf{v}$ bounds the distance between the \ac{ARIS} and the \ac{BS} as follows
    \begin{equation}
        \small u_{u}[\ell] \geq d_{u\text{R}}[\ell],\forall u \in \mathcal{U} \label{d_uR} \text{ and } \forall \ell \in \mathcal{N_T},
    \end{equation}
    \begin{equation}
        \small  v[\ell] \geq d_{\text{RB}}[\ell], \forall \ell \in \mathcal{N_T} \label{d_RB}.
    \end{equation}
Considering the new constraints, the optimization problem is reformulated as 
\begin{subequations} \small
    \begin{alignat}{2}
&\underset{\mathbf{Q,U,v}}{\textrm{minimize}}        &\quad& f_4 (\mathbf{Q},\mathbf{X}) \\
&\textrm{subject to}
&                 & \eqref{UAV-dis}
,\eqref{finalpoint},\eqref{d_uR}, \eqref{d_RB},
\label{TrajPb}
\end{alignat}
\label{TrajPb}
\end{subequations}
where the objective function is expressed as follows
\begin{equation} \small  
\begin{aligned}
\small
    f_5 (\mathbf{Q},\mathbf{X},\mathbf{U},\mathbf{v}) = \sum_{\ell=0}^{N_T} \sum_{u=1}^{U} \sum_{n=1}^{\nc}\delta_{u,n}[\ell] 2 \ x_{u,n}[\ell]~ c_{u,n}[\ell]\\ \small -\delta_{u,n}[\ell] x^2_{u,n}[\ell] \: \left(\frac{a_{u,n}[\ell]}{u_u[\ell]^{\kappa_1}v[\ell]^{\kappa_2}}  + \frac{b_{u,n}[\ell]}{u_u[\ell]^{\frac{\kappa_1}{2}}v[\ell]^{\frac{\kappa_2}{2}}}\right).
    \label{TrajOF}
    \end{aligned}
\end{equation}
Note that the equality at the optimal solution between the original and the new reformulated subproblem is guaranteed since increasing the value of the new slack variables reduces the objective value. However, a minimization problem with a non-convex objective function and/or a non-convex feasible region is generally non-convex, and thus difficult to be solved optimally. Based on these facts, we propose an iterative algorithm to solve problem \eqref{TrajPb} by applying \ac{SCA} \cite{boyd2004convex}.
This approach relies mainly on the first-order Taylor series to linearise the non-convex objective function and constraints. Hence, for given initial points $(\mathbf{U}^{0}, \mathbf{v}^{0})$, which are feasible to \eqref{TrajPb}, we derive the first-order Taylor expansion of \eqref{d_uR}, \eqref{d_RB}, and \eqref{TrajOF} as detailed in Appendix~\ref{AppB}. The approximated problem is written as  
\begin{subequations} \small
\begin{alignat}{3}
&\underset{\mathbf{Q,U,v}}{\textrm{minimize}}        & \quad& \bar{f}_5(\mathbf{Q},\mathbf{X} ,\mathbf{U},\mathbf{V})\label{objFunTraj}\\
&\textrm{subject to}
& \quad &\quad \eqref{UAV-dis},\eqref{finalpoint}, \eqref{approxDUR}, \eqref{approxDRB},
\end{alignat}
\label{ApproxTrajPb}
\end{subequations}
where 
\begin{multline}
\bar{f}_5(\mathbf{Q},\mathbf{X},\mathbf{U},\mathbf{V}) = \sum_{\ell =1 }^{N_T} \sum_{u=1}^{U} \sum_{n=1}^{\nc} \delta_{u,n}[\ell] 2 x_{u,n}[\ell]  c_{u,n}[\ell]\\ - \delta_{u,n}[\ell] x^2_{u,n}[\ell] \left(A_{u,n}[\ell] + B_{u,n}[\ell] \right). 
\end{multline}
After approximation, the problem is convex which can be solved efficiently using convex solvers such as CVX. The \ac{ARIS} trajectory design is detailed in Algorithm~\ref{alg:Traj}. However, it is difficult to generate an initial trajectory that satisfies all constraints. So, we initialize $\mathbf{Q}^0$ as the direct trajectory between the given start and final locations. Based on $\mathbf{Q}^0$, we initialize also the auxiliary variables $(\mathbf{U}^0 , \mathbf{v}^0)$ introduced in the subproblem $(\textrm{P6}^{\prime})$. This guarantees that we have feasible initial points to start the \ac{SCA} algorithm \cite{boyd2004convex}.
Note that the complexity of the proposed algorithm relies on the \ac{SCA} method and can be expressed as $\mathcal{O}\left(N_v^{3.5}\right)$, where $N_v = (3+U)N_T$ is the number of optimization variables.
\begin{algorithm}[t!]
\caption{EMF Exposure-Aware ARIS Trajectory Design}
\label{alg:Traj}
\begin{algorithmic}[1]
\State Initialize $\mathbf{Q}^{(0)}$, $\mathbf{U}^{(0)}, \mathbf{v}^{(0)}, \mathbf{X}^{(0)}$, iteration indices $i=0$, $k=0$, and convergence threshold $\epsilon_5\geq 0$.
\State \textbf{repeat}
\State \quad \textbf{repeat}
\State \qquad Set feasible  $\mathbf{Q}_{\text{feas}}^{(k)}$,$\mathbf{U}_{\text{feas}}^{(k)}$, and $\mathbf{v}_{\text{feas}}^{(k)}$ using $\mathbf{X}^{(i)}$.
\State \qquad Update $\mathbf{Q}^{(k+1)}$ $\mathbf{U}^{(k+1)}$ and $\mathbf{v}^{(k+1)}$ by solving (P6).
\State \qquad Set $k = k + 1$.
\State \quad \textbf{until} convergence.
\State \quad Update $\mathbf{X}^{(i+1)}$ according to Eq.~\eqref{XTraj} for given $\mathbf{Q}^{\ast (i)}$.
\State \quad Set $i = i + 1$.
\State \textbf{until} $\| \mathbf{X}^{(i-1)}-\mathbf{X}^{(i)}\| \leq \epsilon_5$.
\end{algorithmic}
\end{algorithm}
\subsection{Overall Algorithm}
Now that we have studied the optimization of transmit beamforming parameters ($\boldsymbol{\alpha},\boldsymbol{\beta}$), transmit power $\mathbf{P}$, resource allocation vector $\boldsymbol{\delta}$, \ac{ARIS} phase-shift matrix and its trajectory $\mathbf{Q}$.
In this section, we propose the overall \ac{EMF}-aware algorithm and then analyze its complexity. By using the \ac{AO} framework, the overall transmission scheme is detailed in Algorithm \ref{overallAlgo}. The complexity of each subproblem was derived separately. However, the overall complexity of Algorithm~\ref{overallAlgo} mainly centers on two key steps: step (5) the \ac{RIS} phase shifts design, and step (9) the \ac{ARIS} trajectory optimization.

In conclusion, the overall complexity of the proposed solution by reserving the terms with the highest orders is expressed as $\boldsymbol{\mathcal{O}}\left(N_T N^{3.5}+N_v^{3.5}\right)$.
\section{Numerical Results and Discussions}
\label{sec:numericalresults}
\begin{algorithm}[t!]
\caption{Overall EMF Exposure-Aware Algorithm for UL MU-MIMO ARIS-Aided Network}
\label{overallAlgo}
\begin{algorithmic}[1]
\State Initialize $\pv^{(0)}$, iteration index $i=0$ and convergence threshold $\epsilon$
\State \textbf{repeat}
\State \quad \textbf{for} $\ell = 1, 2, \ldots, N_T$ 
\State \qquad Calculate $(\boldsymbol{\alpha}[\ell], \boldsymbol{\beta}[\ell])^{(i+1)}$ using Alg.~\ref{MIMOAlgo}
\State \qquad Calculate $\boldsymbol{\theta}[\ell]^{(i+1)}$ using Alg.~\ref{alg:PhaseShift}
\State \qquad Calculate $\boldsymbol{\delta}[\ell]^{(i+1)}$ using Alg.~\ref{alg:RE}
\State \qquad Calculate $\mathbf{P}[\ell]^{(i+1)}$ using Alg.~\ref{alg:Power}
\State \quad\textbf{end}
\State \quad Calculate $\mathbf{Q}^{(i+1)}$ using Alg.~\ref{alg:Traj}
\State \quad Set $i = i + 1$.
\State \textbf{until}
$\left|\EI\left(\pv^{(i-1)}\right) -\EI\left(\pv^{(i)}\right)\right| \leq \epsilon$.
\end{algorithmic}
\end{algorithm}
\begin{table}[t]
\small
\centering 
\caption{\\ \textsc{Simulation Parameters}}
\renewcommand{\arraystretch}{0.9} 
\setlength{\tabcolsep}{4pt} 
\small 
\begin{tabularx}{\columnwidth}{l X l} 
     \toprule
    \textbf{Parameter} & \textbf{   Description} & \textbf{Value} \\ 
    \hline\hline
    $U$ & Number of users & 8 \\
    $N$ & Total number of channels & 80 \\
    $N_\textrm{c}$ & Number of subcarriers & 80 \\
    $M_\textrm{r}$ & Number of receive antennas & 32 \\
    $M_\textrm{t}$ & Number of transmit antennas & 2 \\
    $h_b$ & BS height & $25$ m \\
    $h_0$ & ARIS height \cite{UAVSpeed} & $100$ m \\
    $\Delta$ & Time slot duration & $15 \, \text{s}$ \\
    $T$ & Total flight duration & $300 \, \text{s}$ \\
    $V_{\text{max}}$ & Maximum velocity \cite{ARISVmax} & $25 \, \text{m/s}$ \\
    $w$ & Bandwidth \cite{NRPPa:20} & $240 \, \text{kHz}$ \\
    $\rho$ & LoS path-loss \cite{3GPP} & $-24.91$ dB \\
    $\rho_1$ & NLoS path-loss \cite{3GPP} & $-19.96$ dB \\
    $P_{\text{max}}$ & Maximum transmit power \cite{NRPPa:20} & $26 \, \text{dBm}$ \\
    $\sigma^2$ & Noise variance & $-174 \, \text{dBm/Hz}$ \\
    $f_\textrm{c}$ & Carrier frequency \cite{UK5GSpectrum} & $700$ MHz \\
    $\kappa$ & Path-loss exponent of UE-BS link \cite{3GPP} & 3.908 \\
    $K_1 = K_2$ & Rician factor of UE-ARIS and ARIS-BS links \cite{ARISVmax} & $3$ dB \\
    $\kappa_1=\kappa_2$ & Path-loss exponent of UE-ARIS and ARIS-BS links \cite{3GPP} & 2.2 \\
    \bottomrule
\end{tabularx}
\label{tab1}
\end{table}
In this section, we present the numerical results obtained by Monte Carlo simulations for our setup. In the evaluation scenario, we assume that the users are uniformly distributed in a circle centred at the origin of the coordinate system, i.e., the \ac{BS} location, with a radius of $100$ m, i.e., the cell coverage. Additionally, we specified the first and final locations of the \ac{ARIS} as, $\mathbf{q}_0 = (-80, 55, 100)$ and $\mathbf{q}_T = (100, 20, 100)$, respectively. Note that the initial and final points of the ARIS trajectory can be selected arbitrarily,  and they can even be the same point, e.g., the same charging station.
\begin{figure*}[t] 
    \begin{minipage}[b]{0.45\textwidth}
        \centering
        \vspace{-1em} 
        \resizebox{\linewidth}{!}{\input{Figures/try}} 
        \caption{Different trajectories of \ac{ARIS} using different benchmark algorithms.}
        \label{Traj}
    \end{minipage}
    \hspace{2em} 
    \begin{minipage}[b]{0.45\textwidth}
        \centering
        \vspace{-1em} 
        \resizebox{\linewidth}{!}{\pgfplotsset{
    every axis/.append style={
        font=\footnotesize,
        line width=1pt,
        legend style={font=\footnotesize, at={(0.5,1.1)}, anchor=south}, 
        legend cell align=left,
    }
}
\begin{tikzpicture}
    \begin{axis}[
        ybar,
        width=0.8\columnwidth, 
        bar width=0.24cm,
        x=0.9cm,
    enlarge x limits=0.1, 
        ymin=0,
        ymax=12,
        symbolic x coords={1, 2, 3, 4, 5, 6, 7, 8},
        xtick=data,
        xtick pos = bottom,
        xlabel={User Index $u$},
        ylabel={Rate [Mbps]},
        legend style={
            at={(0.5,1.04)},
            anchor=south,
            legend columns=-1,
            column sep=1ex
        },
        legend cell align=left,
        ymajorgrids=true,
        grid style=dashed,
        scale only axis,
    ]
    \addplot coordinates {(1,10) (2,9.4) (3,8.5) (4,6.7) (5,4.5) (6,7.6) (7,8.7) (8,3.1)};
    \addplot coordinates {(1,10) (2,9.4) (3,8.5) (4,6.7) (5,4.5) (6,7.6) (7,8.7) (8,3.1)};
    \legend{Achieved rate $r_u$, Minimum required rate $\bar{R}_u$}
    \end{axis}
\end{tikzpicture}} 
        \caption{Comparison of the achieved rate and the minimum required rate for the network users.}
        \label{fig:rateVsU}
    \end{minipage}
\end{figure*}
We consider diverse requirements for the users with the considered rate threshold $\bar{\mathbf{R}} = \left[10,9.4,8.5,6.7,4.5,7.6,8.7,3.1\right]$~Mbps. Here, the path-losses experienced by the \ac{LoS} and \ac{NLoS} links are modeled using 3GPP Urban Micro (UMi) scenario from \cite[Table B.1.2.1-1]{3GPP}. For convenience, the simulation parameters are summarized in Table~\ref{tab1}, unless otherwise specified. The optimization of the convex subproblems in this paper is solved using CVX Toolbox, which is a powerful tool for solving convex problems \cite{CVX}.
We compare the performance of our proposed algorithm with several benchmarks, i.e.,
\begin{itemize}

    \item Without \ac{RIS} \cite{Fabian'spaper}: We rely solely on the direct \ac{NLoS} link between the \ac{UE} and \ac{BS} similar to  \cite{Fabian'spaper} while being adopted to our setup.   {The complexity of this benchmark relies on the transmit power optimization. Hence, its complexity is in the order of $\mathcal{O}\left(\sum_{u=1}^U N_u^2\right)$, where $N_u$ is the number of allocated subcarriers allocated to each user $u \in \mathcal{U}$.}
    \item Fixed Deployed \ac{RIS} \cite{Elzanaty-Emf}: The \ac{RIS} is deployed in a fixed place, i.e., hovering, similar to \cite{Elzanaty-Emf}. For fairness, we extend \cite{Elzanaty-Emf} from \ac{SIMO} to \ac{MIMO} and further optimize the location of the \ac{RIS} to minimize $\EI$ using the divide-and-conquer algorithm.  {The algorithm iteratively divides the search area into subregions, halving the search radius at each step until the optimal ARIS deployment location is identified. Additionally, we account for the time required for the ARIS to travel to and return from the deployment location. The complexity of this benchmark is in the order of $\mathcal{O}\left(N_T N^{3.5}\right)$.} 
    \item \ac{ARIS} with random phases: The proposed algorithm, but with random \ac{RIS} phase shifts, is considered, where all variables are optimized except for the \ac{ARIS} phases.
    \item \ac{ARIS} with zero Phases: The proposed algorithm, but with zero \ac{RIS} phase shifts, is considered, where all variables are optimized except for the \ac{ARIS} phases.
\end{itemize}
\subsection{Impact of \ac{ARIS} Trajectory}
\begin{table*}[t!]
\centering
\caption{\\ \textsc{Average Uplink Exposure ($\EI$) Values for Different ARIS Trajectories}}
\begin{tabularx}{0.89\textwidth}{l *{3}{>{\centering\arraybackslash}X}}
    \toprule
    \textbf{Trajectory} & \textbf{Direct Path} & \textbf{Deployment} & \textbf{Optimized Path} \\
    \midrule
    \hline
    \textbf{$\EI$ [mW/Kg]} & 3.766 & 1.601 & 1.059 \\
    \textbf{Reduction (\%) w.r.t direct path} & -- & 57.48 & 71.88 \\
    \bottomrule
\end{tabularx}
\label{TrajTable}
\end{table*}
\begin{figure}[t]
    \centering
    \pgfplotsset{every axis/.append style={
    font=\footnotesize,
    line width=1pt,
    legend style={fill opacity=0.5, %
        draw opacity=1,  
        text opacity=1,      
font=\footnotesize,at={(0.48,1.09)}, anchor=south, legend columns=1}, 
    legend cell align=left
}}
\pgfplotsset{compat=1.18}

\definecolor{mycolor1}{rgb}{0.49400,0.18400,0.55600}%
\definecolor{mycolor2}{rgb}{0.46600,0.67400,0.18800}%
\definecolor{mycolor3}{rgb}{0.30100,0.74500,0.93300}%
\definecolor{mycolor4}{rgb}{0.63500,0.07800,0.18400}%
\definecolor{mycolor5}{rgb}{0.00000,0.44700,0.74100}%

\begin{tikzpicture}

\begin{axis}[%
xmin=-1e-3,
xmax = 2e-2,
ymin=0,
ymax=1.1,
        yticklabel style={/pgf/number format/precision=1, /pgf/number format/fixed, /pgf/number format/fixed zerofill, /pgf/number format/.cd},
width = 0.9\columnwidth,
ytick ={0.2,0.4,0.6,0.8,1},
grid = major,
xlabel={Exposure [W/Kg]},
ylabel={Cumulative Distribution Function (CDF)},
]

 \addplot[very thick, green,smooth] table {Figures/CDFI.dat};
 \addlegendentry{CDF of the average exposure per user $\EI$ with ARIS}
 \addplot[very thick,red,smooth] table {Figures/CDFMaxE.dat};
\addlegendentry{CDF of Max exposure among users with ARIS}
\addplot[very thick,brown,smooth] table {Figures/CDFMaxWithoutARIS.dat};
  \addlegendentry{CDF of Max exposure among users without ARIS}
 \addplot[dashed, color=mycolor2] coordinates {(0.001209,0) (0.001209,1.1)};

 \addplot[dashed, color=red] coordinates {(0.00258,0) (0.00258,1.1)};

  \addplot[dashed, color=brown] coordinates {(0.0099,0) (0.0099,1.1)};
 
\end{axis}
\end{tikzpicture}%
    \caption{The statistics of maximum and mean exposure among users.}
    \label{fig:CDF}
\end{figure}
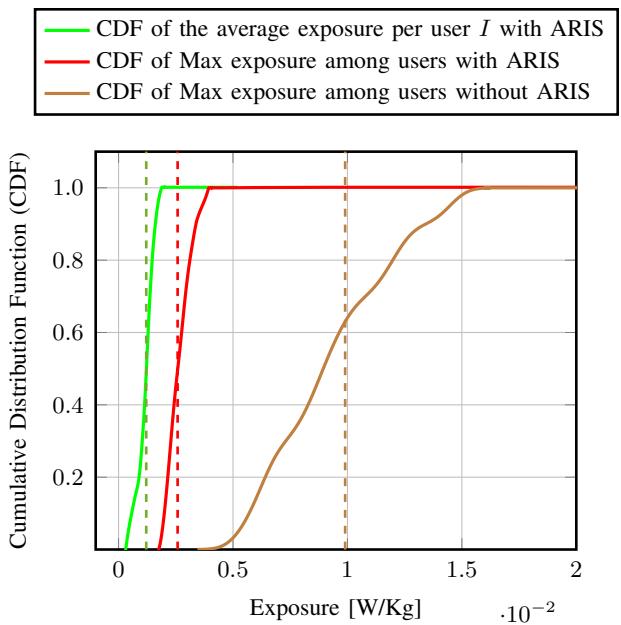
In this section, we show the impact of the \ac{ARIS} trajectory design on the exposure level in Fig.~\ref{Traj} and Table~\ref{TrajTable}. In Fig.~\ref{Traj}, we compare three different trajectories for the \ac{UAV}. One flying path is the proposed designed trajectory to minimize exposure where all the system parameters are optimized, i.e., $\hat{\pv}$. The second trajectory is the direct path between the initial and final locations assuming that all system parameters are optimized except the trajectory $\mathbf{Q}$. The third trajectory considers that the \ac{UAV} is hovering in a specific location for almost all the considered time $T$, except the time required to reach the desired location and the flying time to the final location. A simple interpretation of the ARIS trajectory is that the UAV tends to move closer to the BS and, where possible, to some users. This strategy likely aims to minimize average path-loss, which in turn minimizes the transmit power and, therefore, the exposure. In Table~\ref{TrajTable}, the corresponding exposure for the three considered trajectories is presented. We can see that the proposed trajectory achieves more than 70\% and 30\% reduction in \ac{EMF} exposure compared to the direct trajectory and the one with the \ac{UAV} hovering at a fixed location, respectively. This reduction is achieved while satisfying all rate constraints, as illustrated in Fig.~\ref{fig:rateVsU}. The proposed solution effectively guarantees the fulfillment of QoS requirements while simultaneously minimizing the average uplink exposure.

It is worth noting that the introduction of \ac{ARIS} to the network has a negligible effect on the exposure of neighboring users, as the \ac{ARIS} trajectory is confined within the boundaries of the cell. Moreover, the proposed EMF-aware scheme effectively minimizes the mean exposure across users without increasing individual exposure levels. As shown in Fig.~\ref{fig:CDF}, the maximum exposure without ARIS is significantly higher than that with ARIS. for example, the probability that the maximum exposure with ARIS stays below 10 mW/Kg (which is the mean of the maximum exposure without \ac{ARIS}) is almost 1. Additionally, even in the worst-case scenario, the maximum exposure under ARIS remains below 5 mW/Kg even when the probability is at its highest level, i.e., equal to 1.
\subsection{Impact of the Number of BS Antennas and \ac{RIS} Elements}
\begin{figure}[ht!]
    \centering
    \pgfplotsset{every axis/.append style={
		font=\footnotesize,
		line width=1pt,
		legend style={
  fill opacity=0.5, %
        draw opacity=1,  
        text opacity=1,  
        fill=white,font=\footnotesize, at={(0.63,0.43)}},legend cell align=left},
} %

	\begin{tikzpicture}
\begin{semilogyaxis}[
    xlabel near ticks,
    ylabel near ticks,
    grid=major,
    xtick pos=bottom,
    ytick pos=left,
    xlabel={$\mr$},
    ylabel={$\EI$ [W/Kg]},
    width=0.9\columnwidth,
    xmin=10, xmax=190,
    xtick ={10,70,140,190},    
    ymin=1e-6, 
    ymax=10,
    ytick = {1e0, 1e-2, 1e-4, 1e-6},
    legend entries={Without ARIS \textcolor{blue}{\cite{Fabian'spaper}}, ARIS zero phases, ARIS random phases, ARIS fixed dep. \textcolor{blue}{\cite{Elzanaty-Emf}}, Proposed algorithm},
    ylabel style={font=\normalsize},
    xlabel style={font=\normalsize}
]
    \addplot[brown, mark=o] table {Figures/ExpvsRx/ExpdirectMr.dat};
    \addplot[blue, mark=square] table {Figures/ExpvsRx/ExprndMr.dat};
    \addplot[cyan, mark=oplus*, dashed] table {Figures/ExpvsRx/ExprndMr.dat};
    \addplot[green] table {Figures/ExpvsRx/ExpdepMr.dat};
    \addplot[red, mark=+] table {Figures/ExpvsRx/EhatMr.dat};
    \draw[dashed] (axis cs:10,1.6) -- (axis cs:190,1.6) 
node [pos=0.5, above] {ICNIRP Limit};
\end{semilogyaxis}
\end{tikzpicture}
    \caption{Average uplink exposure $\EI$ as a function of the number of receiving antennas $\mr$ for various methods.}
    \label{E(Mr)}
\end{figure}
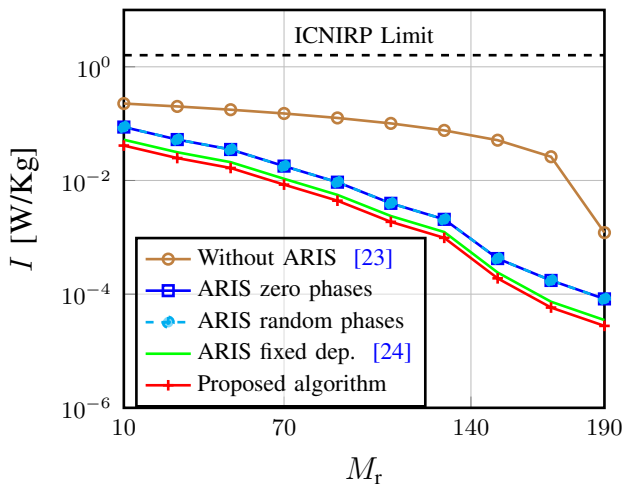
 In Fig.~\ref{E(Mr)}, we evaluate the impact of the number of antennas at the \ac{BS} $\mr$ on the average uplink exposure $\EI$. The results show that the exposure decreases logarithmically when the number of receiving antennas increases.
Increasing the number of antennas at the \ac{BS} improves receive diversity and enhances channel quality. This enables reducing the population's exposure while maintaining the required \ac{QoS}. The proposed algorithm results reveals that using \ac{ARIS} considerably reduces the exposure compared to the fixed \ac{RIS}, which can provide at least 30\% \ac{EMF} exposure reduction for $\mt =32$. Therefore, using \ac{ARIS} can more effectively restrict exposure to meet specific requirements, making it particularly beneficial in environments where limited exposure is crucial, such as schools and hospitals \cite{ULDominance}.
This reduction is attributed to the inherent advantages of \ac{UAV} mobility, which enhances the overall performance of the system. The high mobility allows the \ac{UAV} to dynamically navigate along an optimal trajectory, thereby minimizing the population's exposure to the \ac{RF} radiations.
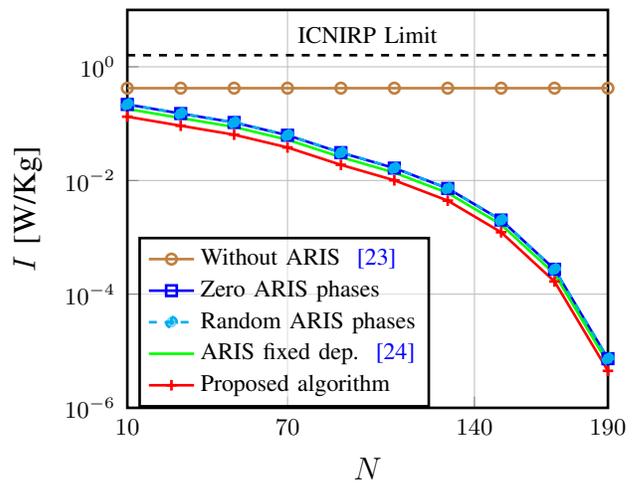
\begin{figure}[t]
    \centering
    \pgfplotsset{every axis/.append style={
		font=\footnotesize,
		line width=1pt,
		legend style={fill opacity=0.5, %
        draw opacity=1,  
        text opacity=1,  
        fill=white,font=\footnotesize, at={(0.63,0.43)}},legend cell align=left},
} %
\pgfplotsset{compat=1.18}
\begin{tikzpicture}
\begin{semilogyaxis}[
     xlabel near ticks,
        ylabel near ticks,
         xtick pos=bottom,
        ytick pos=left,
        ymajorgrids=true,
        xmajorgrids=true,
    grid=major,
    xlabel={$N$},
    ylabel={$\EI$ [W/Kg]},
    width=0.9\columnwidth,
    xmin=10, xmax=190,
    xtick ={10,70,140,190},
    ymin=1e-6, ymax=10,
    ytick={1e-6, 1e-4, 1e-2, 1},
    legend entries={
        {Without ARIS \textcolor{blue}{\cite{Fabian'spaper}}},
        {Zero ARIS phases},
        {Random ARIS phases},
        {ARIS fixed dep. \textcolor{blue}{\cite{Elzanaty-Emf}}},
        {Proposed algorithm}
    },
    ylabel style={font=\normalsize},
    xlabel style={font=\normalsize},
]
 \draw[dashed] (axis cs:10,1.6) -- (axis cs:190,1.6) 
node [pos=0.5, above] {ICNIRP Limit};
\addplot[brown, mark=o] table {Figures/ExpvsRISele/ExpdirectN.dat};
\addplot[blue, mark=square] table {Figures/ExpvsRISele/ExprndN.dat};
\addplot[cyan, mark=oplus*, dashed] table {Figures/ExpvsRISele/ExprndN.dat};
\addplot[green] table {Figures/ExpvsRISele/ExpdepN.dat};
\addplot[red, mark=+] table {Figures/ExpvsRISele/ExphatN.dat};
\end{semilogyaxis}
\end{tikzpicture}
    \caption{Average uplink exposure  $\EI$ as a function of the number of \ac{RIS} elements $N$ for various methods.}
    \label{E(N)}
\end{figure}

Fig.~\ref{E(N)} characterizes the effect of different numbers of \ac{RIS} reflecting elements on the Average uplink exposure  of the network $\EI$. For this evaluation scenario, we assume that the minimum required data rate for all the network users is established at $\bar{R}_{u} = 6$ Mbps, $\forall u \in\mathcal{U}$. Here, we observe that increasing $N$ yields a better performance in terms of exposure reduction. However, for low values of $N$, the reduction is not significant. This can be attributed to the fact that low values of $N$ do not effectively enhance the channel, and therefore do not significantly reduce the transmit power allowing for lower exposure level. In contrast, for $N \geq 100$, the observed reduction is substantial, nearly an order of magnitude lower compared to the scheme where \ac{RIS} is not used. This outcome highlights the effectiveness of \ac{RIS} in mitigating uplink \ac{EMF} exposure when a high number of \ac{RIS} elements are employed. 
\section{Conclusion}
\label{Sec:Conclusion}
In this paper, we introduced a novel architecture incorporating \ac{ARIS} to mitigate the potential risks associated with \ac{EMF} exposure in uplink tranmission of a \ac{MU-MIMO} system. We proposed a two-fold design where we utilized the \ac{ARIS} to establish non-direct \ac{LoS} paths allowing lower transmit power, and to control the transmit beamforming allowing lower induced \ac{SAR} reference. In our scheme, which incorporates \ac{ARIS} mobility, the proposed solution ensures effective reduction of \ac{EMF} exposure through the optimization of the transmit beamforming, \ac{RE} allocation, \ac{RIS} phase-shift vector, the transmit power, and the \ac{ARIS} trajectory while maintaining the required data rate for users. The simulation results confirmed the performance of the proposed algorithm, demonstrating a significant reduction in the \ac{EMF} exposure, i.e., for 6 Mb/s required UL compared to the scheme without RIS. These findings make our proposed architecture highly relevant to the general public, cellular operators, and the research community, as it offers a sustainable solution to mitigate the potential impacts associated with EMF exposure.

\section*{Appendix A\\Power optimization using Lagrangian method}
\label{AppA}
The Lagrangian of the convex optimization problem \eqref{powPb} is
\begin{equation} \small 
\begin{aligned}
\small 
&\mathcal{L}\left(\mathbf{p}_u,\mu,\lambda\right) = \sum_{n=1}^{\nc} \delta_{u,n} \Big( \overline{\mathrm{SAR}}_{u,n}~ p_{u,n}\\ 
&+ \mu \left(\bar{R}_{u} - r_{u} (\pv)\right) 
+ \lambda \left( \sum_{n=1}^{\nc} \delta_{u, n} \ p_{u,n} - P_{\text{max}}\right) \Big),
\end{aligned}
\label{lagrange}
\end{equation}
 where $\mu$ and $\lambda$ denote the Lagrangian multipliers associated with \eqref{QoSConstraint} and \eqref{PowerConstraint}, respectively. Noting that the Lagrangian function is convex w.r.t $\mathbf{p}_u$, the optimal solution for this problem needs to satisfy the following KKT conditions
\begin{align}
\begin{cases}
\small
\frac{\partial{\mathcal{L}}(\mathbf{p}_u,\mu,\lambda)}{\partial{\mathbf{p}}_u} = \mathbf{0}, \\
\small
\mu \left(\bar{R}_{u}- r_{u}(\pv_{u,n})\right) = 0, & \mu \geq 0, \\
\small
\lambda ( \sum_{n=1}^{\nc} \delta_{u, n} \ p_{u,n}-P_{\text{max}}
) = 0,&  \lambda \geq 0.
\label{SysEqPowOpt}
\end{cases}
\end{align}
Taking the derivative of $\mathcal{L}$ \ac{w.r.t} $p_{u,n}$, we obtain
\begin{equation} \small
\begin{aligned}
\small
    \frac{\partial \mathcal{L}(\mathbf{p}_u,\mu,\lambda)}{\partial p_{u,n}} = \delta_{u,n} \left(\frac{w \overline{\mathrm{SAR}}_{u,n} \mu \gamma(\pv_{u,n})}{\mathrm{ln}(2)(\sigma^2+p_{u,n} 
\gamma(\pv_{u,n}))} + \lambda\right).
    \label{lagrange}
    \end{aligned}
\end{equation}
Then, based on \eqref{lagrange}, we can derive the optimal power as \eqref{optPower}. 
On the other hand, the optimal $\mu^\ast$ and $\lambda^\ast$ can be numerically obtained by solving 
\eqref{SysEqPowOpt}.
\section*{Appendix B\\SCA for ARIS trajectory design}
\label{AppB}
The first-order Taylor expansions of \eqref{TrajOF}, \eqref{d_uR}, and \eqref{d_RB} at given feasible points ($\mathbf{u}_{u}^{0}, \mathbf{v}^{0}$) can be respectively expressed as
\begin{equation} \small
\begin{aligned}
\small
   x^2_{u,n}[\ell] \left( \frac{a_{u,n}[\ell]}{u_u[\ell]^{\kappa_1}v[\ell]^{\kappa_2}}  + \frac{b_{u,n}[\ell]}{u_u[\ell]^{\frac{\kappa_1}{2}}v[\ell]^{\frac{\kappa_2}{2}}} \right)\\\small \leq x^2_{u,n}[\ell] \left(A_{u,n}[\ell] + B_{u,n}[\ell] \right),
   \end{aligned}
\end{equation}
\begin{align}
\centering
\small
     d_{u\text{R}}^2[\ell]+(u_{u}^{0}[\ell])^2   - 2~u_{u}^{0}[\ell]~ u_{u}[\ell]\leq 0, &\small ~\forall u \in \mathcal{U} , \forall \ell \in \mathcal{N_T}, \label{approxDUR} \\
     \small
     d_{\text{RB}}^2[\ell]+(v^{0}[\ell])^2 - 2~v^{0}[\ell]~ v[\ell] \leq 0, &~ \small\forall \ell \in \mathcal{N_T},\label{approxDRB}
\end{align}
where 
\begin{equation} \small
\begin{aligned}
&\small A_{u,n}[\ell] =  \, a_{u,n}[\ell] \Bigg( 
\frac{1}{(u_{u}^{0}[\ell])^{\kappa_1} + (v^{0}[\ell])^{\kappa_2}}\\&\small +
\frac{\kappa_1 (u_u[\ell] - u_{u}^{0}[\ell])}{(u_{u}^{0}[\ell])^{\kappa_1 + 1} + v[\ell]^{\kappa_2}} 
 - \frac{\kappa_2 (v[\ell] - v^{0}[\ell])}{u_{u}[\ell]^{\kappa_1} + (v^{0}[\ell])^{\kappa_2 + 1}}
\Bigg),
\end{aligned}
\end{equation}
and
\begin{equation} \small
\begin{aligned}
&\small B_{u,n}[\ell] =  \, b_{u,n}[\ell] \Bigg(
\frac{1}{(u_{u}^{0}[\ell])^{\frac{\kappa_1}{2}} + (v^{0}[\ell])^{\frac{\kappa_2}{2}}}\\ & \small
+ \frac{-\frac{\kappa_1}{2} (u_u[\ell] - u_{u}^{0}[\ell])}{(u_{u}^{0}[\ell])^{\frac{\kappa_1}{2} + 1} + v[\ell]^{\frac{\kappa_2}{2}}} 
 - \frac{\frac{\kappa_2}{2} (v[\ell] - v^{0}[\ell])}{u_{u}[\ell]^{\frac{\kappa_1}{2}} + (v^{0}[\ell])^{\frac{\kappa_2}{2} + 1}}
\Bigg).
\end{aligned}
\end{equation}

 As such, \eqref{TrajPb} can be approximated as \eqref{ApproxTrajPb}.
\bibliographystyle{IEEEtran}
\enlargethispage{0cm}
\bibliography{IEEEabrv}
\end{document}